\documentclass[prd,preprintnumbers,showkeys,nofootinbib,11pt]{article}
\usepackage{epsfig}

\usepackage{bm, graphicx,amsmath,amssymb,epsf}
\usepackage{appendix}

\usepackage{verbatim}

\usepackage{subfigure}

\usepackage{graphicx}
\newcommand{\beq}{\begin{equation}}
\newcommand{\eeq}{\end{equation}}
\newcommand{\beqn}{\begin{eqnarray}}
\newcommand{\eeqn}{\end{eqnarray}}
\newcommand{\beqs}{\begin{eqnarray*}}
\newcommand{\eeqs}{\end{eqnarray*}}

\begin{document}

\title{\large \bf BFKL approach   and six-particle MHV amplitude in $\mathcal{N}=4$ super Yang-Mills}
\author{\large  L.~N. Lipatov$^{1,2}$ and
A.~Prygarin$^{1}$ \bigskip \\
{\it
$^1$~II. Institute of  Theoretical Physics, Hamburg University, Germany} \\
{\it  $^2$~St. Petersburg Nuclear Physics Institute, Russia}}

\maketitle

\vspace{-9cm}
\begin{flushright}
DESY-10-173
\end{flushright}
\vspace{8cm}

\abstract{
We consider the planar  MHV amplitude in $\mathcal{N}=4$ supersymmetric Yang-Mills theory for $2 \to 4$ particle scattering  at two and three loops in the Regge kinematics. We perform an analytic continuation of two-loop result for the remainder function found by Goncharov, Spradlin,  Vergu and  Volovich to the physical region, where the remainder function does not vanish in the Regge limit.  After the  continuation both the leading and the subleading  in the logarithm of the energy terms are extracted and analyzed. Using this result we calculate the next-to-leading corrections to the impact factors required in the BFKL approach.  The BFKL technique was used to find the leading imaginary and real parts of  the remainder function at three loops.
}

\newpage



\section{Introduction}
Recently we have witnessed  revolutionary developments in studying scattering amplitudes in supersymmetric theories.
The present progress is traced back to the work of Parke and Taylor~\cite{Parke:1986gb}, who showed that the tree-level gluon scattering amplitude can be written in a very compact form for some particular helicities of the external particles, namely the maximally helicity violating~(MHV) amplitude. The simplicity of the Parke-Taylor tree-level formula raised a hope that the quantum corrections could be also compactly encoded in the MHV gluon amplitudes.
A great effort in this direction led to formulation by Anastasiou, Bern, Dixon and Kosower~(ABDK)~\cite{Anastasiou:2003kj} and then by Bern, Dixon and Smirnov~(BDS)~\cite{BDS} ansatz for multi-loop planar gluon  MHV amplitude in $\mathcal{N}=4$ super Yang-Mills theory.

 The BDS formula was tested by Alday and Maldacena~\cite{Alday:2007he} from strong coupling side using conjectured AdS/CFT correspondence in the limit of large number of external legs. They argued that the BDS ansatz is probably to be violated  starting at six external gluons. This violation was established by Bartels, Sabio Vera and one of the authors~(BLS)~\cite{BLS1} analyzing the analytic structure of the BDS amplitude. It was shown that the BDS ansatz for six-particle amplitude at two loops is not compatible with the Steinmann relations~\cite{Steinmann}, that impose the absence of the simultaneous singularities in the overlapping channels. They also showed~\cite{BLS2} that the BDS violating piece originates from the so-called Mandelstam cuts, which are the moving Regge singularities in the complex angular momenta plane. The BDS violating term in the multi-Regge kinematics  was explicitly calculated~\cite{BLS2} with logarithmic accuracy in the physical region, where
it gives a non-vanishing and pure imaginary contribution. We call this region the Mandelstam region~(channel). The BDS violating term for the six-point planar MHV amplitude was found using the BFKL approach~\cite{BFKL} and for an arbitrary number of external gluons it contains contributions of Mandelstam cuts constructed from an arbitrary number of reggeized gluons for the Bartels-Kwiecinski-Praszalowicz~(BKP) state~\cite{Kwiecinski:1980wb,Bartels:1993ih} with the local Hamiltonian of an integrable open Heisenberg spin chain~\cite{Lipatov:2009nt}. Other interesting limits of   MHV amplitudes were studied in the Regge kinematics by Brower, Nastase, Schnitzer and  Chung-I Tan~\cite{Brower:2008ia}.

Drummond, Henn, Korchemsky and Sokatchev~\cite{Drummond:2007au} analyzed the conformal properties of polygon Wilson loops in $\mathcal{N}=4$ SYM and showed that anomalous conformal Ward identities  uniquely
fix the form of the all-loop $4$- and  $5$-point amplitudes, so that any relative correction to the BDS ansatz starting at six external particles should be a  function of conformal invariants (cross ratios of dual coordinates).
  The relative correction to the BDS formula was named the remainder function $R^{(L)}_{n}$ for an amplitude with $L$ loops and $n$ external legs, and the first non-trivial remainder function is  $R^{(2)}_{6}$.

 It was suggested~\cite{Alday:2007hr,DKS,Brandhuber:2007yx,Berkovits:2008ic,Drummond:2007bm,Drumlast} that $R^{(L)}_{n}$ can be obtained from the expectation value of the light-like  polygonal  Wilson loops.   Del Duca, Duhr and Smirnov~\cite{DelDuca:2009au,DelDuca:2010zg} expressed $R^{(2)}_{6}$ in terms of   generalized polylogarithms, which was greatly simplified by Goncharov, Spradlin, Vergu and Volovich~(GSVV) \cite{Goncharov:2010jf}, and then written only in terms of $\text{Li}_k$ functions   with arguments depending on three dual conformal cross ratios.

The two-loop remainder function $R^{(2)}_{8}$  for the scattering of eight gluons  was calculated  by Del Duca, Duhr and Smirnov~\cite{DelDuca:2010zp} and its diagrammatic structure was analyzed by Alday~\cite{Alday:2010jz}. The form of $R^{(2)}_{8}$ is remarkably simple and it is constructed  only of  a product of some logarithms plus a constant term.

Earlier we  performed~\cite{Lipatov:2010qg} an analytic continuation of the GSVV formula to a physical region considered in refs.~\cite{BLS1,BLS2}. The  continuation showed a full agreement between the BLS formula and the Wilson loop calculations at the leading logarithmic level and allowed to extract the terms subleading in the logarithm of the energy.  Numerically, an agreement between the two approaches was demonstrated by Schabinger~\cite{Schabinger:2009bb}. The analytic continuation to the mentioned above physical region in the regime of the strong coupling constant was performed by Bartels, Kotanski and Schomerus~\cite{Bartels:2010ej}. They found the leading singularity, which governs the high energy behavior of the scattering amplitude, so-called reggeon intercept. At weak coupling constant the corresponding intercept is determined by the BFKL equation~\cite{BLS2}.

In this study we present some details of the analytic continuation performed by the authors in ref.~\cite{Lipatov:2010qg}. Based  on the obtained  result we calculate the next-to-leading~(NLO) impact factors for the color octet states in the BFKL approach.
In the BFKL technique we also find  the three-loop contribution to the remainder function of  planar six-point MHV amplitude in the leading logarithmic approximation~(LLA) as well as the real part of the subleading corrections in the next-to-leading logarithmic approximation.

\section{BFKL approach}

\hspace{0.6cm}In this section we briefly outline the results of the BFKL approach  to the planar MHV amplitudes in $\mathcal{N}=4$ SYM.

\begin{figure}[htbp]
	\begin{center}
	\subfigure[$u_1=e^{i0}|u_1|$~($s,s_2,s_1,s_3,s_{012},s_{123}>0$)]{
	\epsfig{figure=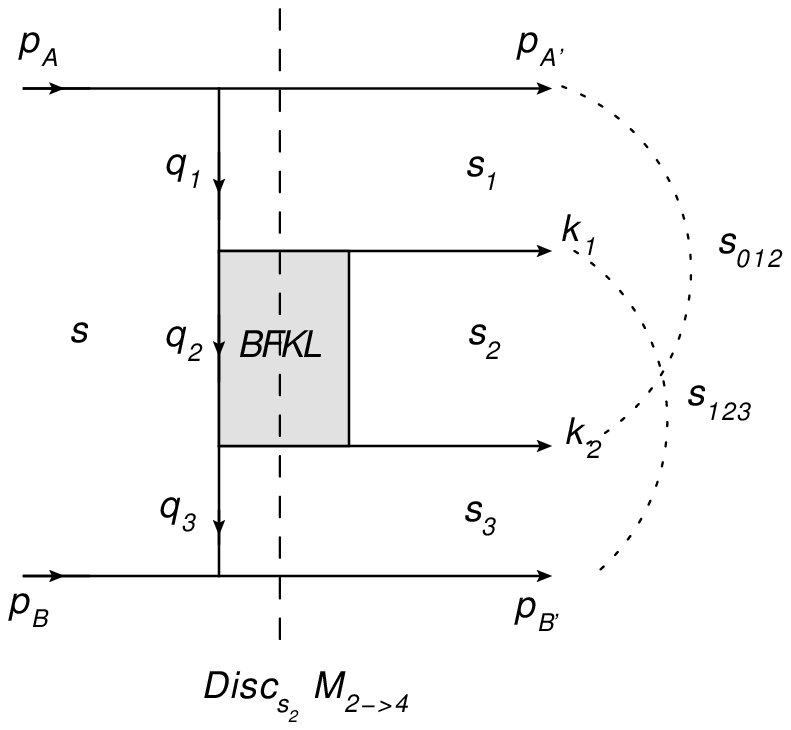,width=70mm}}
	\subfigure[$u_1=e^{-i2\pi}|u_1|$~($s,s_2>0;\,s_1,s_3,s_{012},s_{123}<0$)]{
	\epsfig{figure=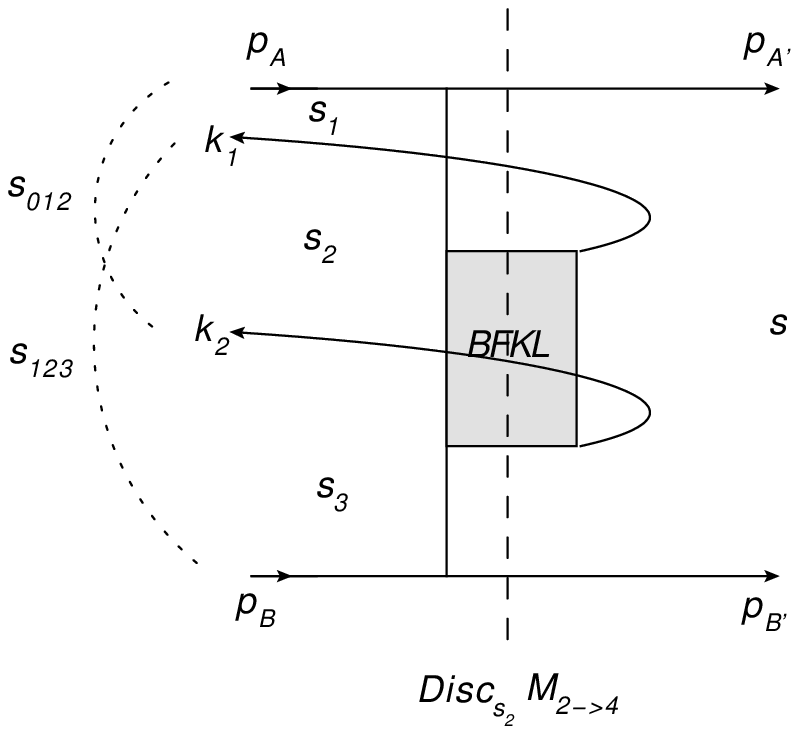,width=70mm}}
		\end{center}
	\caption{ The BDS violating contribution appears in the region $s,s_2>0;\,s_1,s_3<0$.   }
	\label{fig:6pointI}
\end{figure}

 We consider the six-point MHV amplitude
for production of two gluons with momenta $k_1$ and $k_2$ in small angle scattering of the particles with momenta $p_A$ and $p_B$ as depicted in Fig.~\ref{fig:6pointI}.

All energy invariants are shown in  Fig.~\ref{fig:6pointI} and are related to the dual conformal cross ratios by
\begin{eqnarray}\label{crossinv}
 u_1=\frac{s\;s_2}{s_{012}\;s_{123}},\;\;\; u_2=\frac{s_1\;t_3}{s_{012}\;t_{2}},\;\;\; u_3=\frac{s_3\;t_1}{s_{123}\;t_2}.
\end{eqnarray}

 The multi-Regge kinematics is equivalent to having
$s\gg s_{012}, s_{123} \gg s_1, s_2, s_3 \gg t_1, t_2, t_3$,
which in the terms of the cross ratios reads

 \begin{eqnarray}\label{MRK}
1-u_1\to +0,\;\; u_{2}\to +0,\;\;u_{3}\to + 0,\;\; \frac{u_2}{1-u_1}\simeq \mathcal{O}(1)
,\;\; \frac{u_3}{1-u_1}\simeq \mathcal{O}(1).
\end{eqnarray}

In this kinematics the remainder function of the MHV amplitude goes to zero in direct channel in Fig.~\ref{fig:6pointI}a,  while in the Mandelstam channel Fig.~\ref{fig:6pointI}b grows with $\ln s_2$ and becomes pure imaginary. In the Mandelstam channel the gluon momenta $k_1$ and $k_2$ are flipped and the cross ratio $u_1$ possesses a phase
\beqn
{u_1}_b={u_1}_a e^{-i2\pi},
\eeqn
leading to necessity of an analytic continuation of $R_6$.
It was demonstrated by Bartels, Sabio Vera and one of the authors~\cite{BLS1,BLS2} that  the BDS violating piece comes from the Mandelstam cut   state propagating in the crossing channel between the produced particles $k_1$ and $k_2$, and denoted by  the dark box in Fig.~\ref{fig:6pointI}. In $\mathcal{N}=4$ SYM (as well in QCD) for a large number of colors  this state
is  described by the color octet BFKL evolution equation~\cite{BFKL}. The BFKL equation can be formulated as the Schr\"odinger equation  with a Hamiltonian equivalent to that of a completely  integrable open Heisenberg spin chain model~\cite{Lipatov:2009nt}, which made it possible to solve it analytically~\cite{BLS2}.
In the direct channel of the multi-Regge kinematics~(see Fig.~\ref{fig:6pointI}a)  given by   Eq.~\ref{MRK} the remainder function vanishes due to the Mandelstam cancellation of cut contributions as was shown in ref.~\cite{BLS1}.

 The BDS violating piece in the Mandelstam channel is given by~\cite{BLS2}
\beq
M_{2\rightarrow 4}=M^{BDS}_{2\rightarrow 4}\,(1+i\Delta _{2\rightarrow 4}),
\label{corLLA}
\eeq
where $M^{BDS}_{2\rightarrow 4}$ is the BDS amplitude~\cite{BDS} and the correction $\Delta _{2\rightarrow 4}$ was calculated
in all orders with a leading logarithmic accuracy using the solution to the octet BFKL equation. The all-orders LLA expression  for $\Delta_{2\to 4}$ reads
\beqn\label{LLA}
&& \Delta _{2\rightarrow 4}=\frac{a}{2}\, \sum _{n=-\infty}^\infty (-1)^n
\int _{-\infty}^\infty \frac{d\nu }{\nu ^2+\frac{n^2}{4}}\,
\left(\frac{q_3^*k^*_1}{k^*_2q_1^*}\right)^{i\nu -\frac{n}{2}}\,
\left(\frac{q_3k_1}{k_2q_1}\right)^{i\nu +\frac{n}{2}}\,
\left(s_2^ {\omega (\nu , n)}-1\right)\,\\
&&
\simeq \frac{a}{2}\, \sum _{n=-\infty}^\infty (-1)^n
\int _{-\infty}^\infty \frac{d\nu }{\nu ^2+\frac{n^2}{4}}\,
\left(\frac{q_3^*k^*_1}{k^*_2q_1^*}\right)^{i\nu -\frac{n}{2}}\,
\left(\frac{q_3k_1}{k_2q_1}\right)^{i\nu +\frac{n}{2}}\,
\left((1-u_1)^ {-\omega (\nu , n)}-1\right) \nonumber
\eeqn
Here $k_1,k_2$ are transverse components of produced gluon momenta,
$q_1,q_2,q_3$ are the momenta of reggeons in the corresponding
crossing channels and
\beq\label{eigen}
\omega (\nu , n)=-a E_{\nu,n}.
\eeq
 The perturbation theory parameter $a$ and the eigenvalue of the color octet BFKL $E_{\nu,n}$ are given by
\beqn
a=\frac{g^2 N_c}{8\pi^2}\left(4\pi e^{-\gamma}\right)^\epsilon \Rightarrow
\frac{\alpha_s N_c}{2\pi}
\eeqn
and
\beqn\label{Enun}
E_{\nu,n}=-\frac{1}{2}\frac{|n|}{\nu^2+\frac{n^2}{4}}+\psi\left(1+i\nu+\frac{|n|}{2}\right)+\psi\left(1-i\nu+\frac{|n|}{2}\right)-2\psi(1),
\eeqn
where $\psi(z)=\Gamma'(z)/\Gamma(z)$, $\gamma$ is the Euler constant $\gamma=-\psi(1)$ and the dimensional regularization parameter $\epsilon$ is defined by $d=4-2\epsilon$.

The second line of Eq.~\ref{LLA} follows from the fact that in the Regge kinematics the energy invariant $s_2$ is related to the cross ratio $u_1$ by
\beqn\label{u1s2}
1-u_1 \simeq \frac{(\mathbf{k}_1+\mathbf{k}_2)^2}{s_2}=\frac{s_0}{s_2}.
\eeqn
This way we set $s_0=(\mathbf{k}_1+\mathbf{k}_2)^2$ to be an energy scale, which becomes relevant only beyond leading logarithmic approximation. The choice of the energy scale $s_0$ is natural in the Regge kinematics because it reflects the smallness of the transverse components with respect to the longitudinal components of the particle  momenta. The expression in Eq.~\ref{LLA} is a function of the dual conformal cross ratios $u_i$ as discussed in section~\ref{sec:NLO}.
The BDS violating piece at two loops  found in ref.~\cite{BLS2}  can be written in terms of the reduced cross ratios as
\begin{eqnarray}\label{KblsR6}
 && a^2 R^{(2)\; LLA}=i\Delta _{2\rightarrow 4}=-i a^2 \frac{\pi}{2}\ln s_2 \ln \left(\frac{|\bold{k}_2+\bold{k}_1|^2 |\bold{q}_2|^2}{|\bold{k}_2|^2|\bold{q}_1|^2}\right)
\ln \left(\frac{|\bold{k}_2+\bold{k}_1|^2 |\bold{q}_2|^2}{|\bold{k}_1|^2|\bold{q}_3|^2}\right)  \\
&&\simeq i a^2\frac{\pi}{2}\ln(1-u_1)\ln \tilde{u}_2 \ln \tilde{u}_3
\nonumber
\end{eqnarray}
using Eq.~\ref{u1s2} and the fact that the reduced cross ratios
\beqn\label{redcross}
\tilde{u}_2=\frac{u_2}{1-u_1}, \;\; \tilde{u}_3=\frac{u_3}{1-u_1}
\eeqn

 in the multi-Regge kinematics are given by

\begin{eqnarray}\label{redcrosstrans}
 \tilde{u}_2 \simeq \frac{|\bold{k}_2|^2|\bold{q}_1|^2}{|\bold{k}_2+\bold{k}_1|^2 |\bold{q}_2|^2} , \;\;\;\tilde{u}_3 \simeq\frac{|\bold{k}_1|^2|\bold{q}_3|^2}{|\bold{k}_2+\bold{k}_1|^2 |\bold{q}_2|^2}.
\end{eqnarray}

Surprisingly, the expression in Eq.~\ref{KblsR6} can be obtained from the BDS formula using only general analytic properties of the scattering amplitudes and the  factorization hypothesis (proposed by  Alday and Maldacena~\cite{Alday:2007hr}) as it was  shown by one of the authors~\cite{Lipatov:2010qf}. In this technique it is enough to know the form of the BDS amplitude at one loop to obtain the leading logarithmic imaginary term  at two loops. Unfortunately, for the three loops the knowledge of only the BDS formula is not enough and some extra information is to be included in the analysis. This may come from the full analytic form of the remainder function at two loops $R_{6}^{(2)}$.
The function $R_{6}^{(2)}$ was calculated by Drummond, Henn, Korchemsky and Sokatchev~\cite{Drummond:2008aq}  using  the duality between the light-like Wilson loops and the MHV amplitudes, and then greatly simplified by Goncharov, Spradlin, Vergu and Volovich~\cite{Goncharov:2010jf} using the integral representation of Del Duca, Duhr and Smirnov~\cite{DelDuca:2009au,DelDuca:2010zg}. In the next section we perform the analytic continuation of the two-loop remainder function calculated by Goncharov, Spradlin, Vergu and Volovich to the region where $u_1=|u_1|e^{-i2\pi}$, which corresponds to the Mandelstam channel in Fig.~\ref{fig:6pointI}b in the multi-Regge kinematics.

\section{Analytic continuation}

In this section we discuss some details of the analytic continuation to the Mandelstam channel in the Regge kinematics. We also show how the kinematics determines the physical region of the cross ratios and establish the match between our picture and the one drawn by Alday, Gaiotto and Maldacena~\cite{Alday:2009dv}.

The result of Goncharov, Spradlin, Vergu and Volovich~\cite{Goncharov:2010jf} for the two-loop remainder function reads~\footnote{When the present manuscript was already at the last stage of the preparation, a new version of ref.~\cite{Goncharov:2010jf} appeared. The non-analytic term $\chi$ was eliminated in the new version. This fact does not affect our result so that here we use the initial version of the GSVV formula.}
\begin{eqnarray} \label{R6}
R^{(2)}_6(u_1,u_2,u_3) = \sum_{i=1}^3 \left( L_4(x^+_i, x^-_i) -
\frac{1}{2} \text{Li}_4(1 - 1/u_i)\right) \cr
- \frac{1}{8} \left( \sum_{i=1}^3 \text{Li}_2(1 - 1/u_i) \right)^2
+ \frac{J^4}{24} + \chi \frac{\pi^2}{12} \left(  J^2 +  \zeta(2)\right),
\end{eqnarray}
where
\begin{equation}\label{Xpm}
x^\pm_i = u_i x^\pm, \qquad
x^\pm = \frac{u_1+u_2+u_3-1 \pm \sqrt{\Delta}}{2 u_1 u_2 u_3},
\end{equation}
and $\Delta = (u_1+u_2+u_3-1)^2 - 4 u_1u_2u_3$.

The function $L_4(x^+, x^-)$ is defined by

\begin{equation}
\label{eq:bwrz}
L_4(x^+, x^-) = \sum_{m=0}^3
\frac{(-1)^m}{(2m)!!} \log(x^+ x^-)^m
(\ell_{4-m}(x^+) + \ell_{4-m}(x^-))
+ \frac{1}{8!!} \log(x^+ x^-)^4,
\end{equation}
together with
\begin{equation}
\ell_n(x) = \frac{1}{2} \left( \text{Li}_n(x) - (-1)^n \text{Li}_n(1/x) \right),
\end{equation}
as well as the quantities
\begin{equation}\label{J}
J = \sum_{i=1}^3 (\ell_1(x^+_i) - \ell_1(x^-_i)),
\end{equation}
and
\begin{equation}\label{chi}
\begin{aligned}
\chi &= \begin{cases}
-2 & \Delta > 0 ~{\rm and}~ u_1+u_2+u_3>1,\cr
+1 & {\rm otherwise}.
\end{cases}
\end{aligned}
\end{equation}

The result of the analytic continuation to the Mandelstam channel illustrated in Fig.~\ref{fig:6pointI}b, where  $u_1=|u_1|e^{-i2\pi}$, was presented by the us in ref.~\cite{Lipatov:2010qg} and reads (see Appendices \ref{app:NLLA} and \ref{app:complex} for more details)

 \begin{eqnarray}\label{KR62result}
&&R^{(2)\; LLA+NLLA}_6(|u_1|e^{-i2\pi},|z|^2(1-u_1),|1-z|^2(1-u_1))\simeq \frac{i\pi}{2}\ln(1-u_1)\ln |z|^2 \ln |1-z|^2  \nonumber\\
&&+\frac{i\pi}{2} \ln \left(|z|^2|1-z|^2\right) \left(\ln z \ln (1-z)+\ln z^* \ln (1-z^*) -2\zeta_2\right)   \nonumber\\
&& +\frac{i\pi}{2} \ln \frac{|1-z|^2}{|z|^2}\left(\text{Li}_2(z)+\text{Li}_2(z^*)-\text{Li}_2(1-z)-\text{Li}_2(1-z^*)\right)
\nonumber \\
&& +i2\pi \left(\text{Li}_3(z)+\text{Li}_3(z^*)+\text{Li}_3(1-z)+\text{Li}_3(1-z^*)-2\zeta_3\right).
\end{eqnarray}
In Eq.~\ref{KR62result} we introduced complex variables
\begin{eqnarray}\label{Kcomplex}
z=\sqrt{\frac{u_2}{1-u_1}}e^{i\phi_2}=\sqrt{\tilde{u}_2}e^{i\phi_2}
,\;\;\;1-z=\sqrt{\frac{u_3}{1-u_1}}e^{-i\phi_3}=\sqrt{\tilde{u}_3}e^{-i\phi_3}
\end{eqnarray}
to remove some square roots in the arguments of the polylogarithms~(see Eq.~\ref{Xpm}).
The phases $\phi_2$ and $\phi_3$ can be easily expressed in terms of the cross ratios $u_i$ and have a meaning of the angles of the "unitarity" triangle illustrated in Fig.~\ref{fig:triangle}. More details on this parametrization  are presented in the appendix~\ref{app:complex}.

\begin{figure}[htbp]
	\begin{center}
		\epsfig{figure=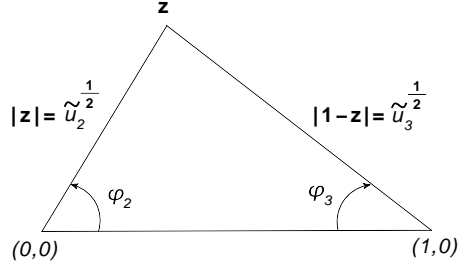,width=60mm}
	\end{center}
	\caption{The "unitarity" triangle.   }
	\label{fig:triangle}
\end{figure}

The first term on RHS of Eq.~\ref{KR62result} reproduces the leading logarithm term found by Bartels, Sabio Vera and one of the authors~\cite{BLS2} in the BFKL approach as explained below. It is easy to see from Eq.~\ref{Kcomplex}  that
\beqn
\ln |z|^2 \ln|1-z|^2=\ln \tilde{u}_2 \ln \tilde{u}_3
\eeqn
 and thus the first term on RHS  in  Eq.~\ref{KR62result} equals to $R^{(2)\;LLA}$ in Eq.~\ref{KblsR6}.

 Other terms correspond to  the next-to-leading logarithmic approximation ~(NLLA) and they present a new result, which is yet to be calculated using the BFKL technique. This analysis shows an agreement between the conjectured duality between the light-like Wilson loops and the MHV amplitudes, and the BFKL approach at the leading logarithmic level.

The remainder function of Eq.~\ref{KR62result} in this channel in the multi-Regge limit is pure imaginary
 and symmetric under the substitution $z \leftrightarrow 1-z$, which corresponds to the target-projectile symmetry in Fig.~\ref{fig:6pointI}.   Eq.~\ref{KR62result}  vanishes for $z\to 1$ or $z\to 0$, when the momentum of one of the produced particles $k_i$ in Fig.~\ref{fig:6pointI} goes to zero, in an accordance to the expectation that in the collinear limit the six-point amplitude  reduces to the five-point amplitude.

Another useful form of the remainder function in Eq.~\ref{KR62result} can be written as
\begin{eqnarray}\label{R62w}
 &&R^{(2)\;LLA+NLLA}_6\left(|u_1|e^{-i2\pi},\frac{1}{|1+w|^2},\frac{|w|^2}{|1+w|^2}\right)\simeq\frac{i\pi}{2}\ln(1-u_1) \ln |1+w|^2 \ln \left|1+\frac{1}{w}\right|^2 \hspace{0.5cm} \\
&& +\frac{i\pi}{2} \ln |w|^2 \ln^2|1+w|^2-\frac{i\pi}{3}\ln^3 |1+w|^2+ i\pi \ln |w|^2 \left( \text{Li}_2 (-w) +\text{Li}_2 (-w^*)\right)
\nonumber \\
&&-i2\pi  \left( \text{Li}_3 (-w) +\text{Li}_3 (-w^*)\right),
\nonumber
\end{eqnarray}
where the complex variable $w$ is expressed in terms of the reduced cross ratios of Eq.~\ref{redcross} as
 \beqn\label{wtext}
w=\frac{1-z}{z}=\frac{B^+}{\tilde{u}_2},\;\;\; w^*=\frac{1-z^*}{z^*}=\frac{B^-}{\tilde{u}_2}
 \eeqn
for $B^{\pm}$ defined in Eq.~\ref{Bpm} by
\begin{eqnarray}\label{Bpmtext}
B^{\pm}=\frac{1-\tilde{u}_2-\tilde{u}_3\pm \sqrt{(1-\tilde{u}_2-\tilde{u}_3)^2-4\tilde{u}_2\tilde{u}_3}}{2}.
\end{eqnarray}

The complete discussion on the details of the analytic continuation and the $z$ and $w$ representations of the remainder function is presented in the appendices~A-E, and here we only want  to emphasize some important points. The analytic continuation was performed  under an assumption that  $u_1+u_2+u_3<1$ to avoid a difficulty related to non-analyticity of $\chi$ in  Eq.~\ref{chi}. We made sure that after the continuation the remainder function does not have any singularities on the border of this region and therefore it is valid in the whole  physical region.

\begin{figure}[tbp]
	\begin{center}
		\epsfig{figure=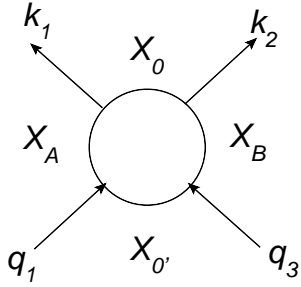,width=40mm}
	\end{center}
	\caption{The dual coordinates of the transverse momenta.   }
	\label{fig:dual}
\end{figure}

The variables $\tilde{u}_2$ and $\tilde{u}_3$ in Eq.~\ref{redcross} are also cross ratios in the transverse momentum space as can be seen from  Eq.~\ref{redcrosstrans} defining the dual coordinates in the transverse momenta space as illustrated in Fig.~\ref{fig:dual}.

In terms of the dual coordinates the reduced cross ratios in Eq.~\ref{redcrosstrans} read
\begin{eqnarray}\label{crossX}
\tilde{u}_2=\frac{|x_{0B}|^2|x_{0'A}|^2}{|x_{AB}|^2|x_{00'}|^2}, \;\;\;
 \tilde{u}_3=\frac{|x_{0A}|^2|x_{0'B}|^2}{|x_{AB}|^2|x_{00'}|^2}.
\end{eqnarray}

\begin{figure}[htbp]
	\begin{center}
		\epsfig{figure=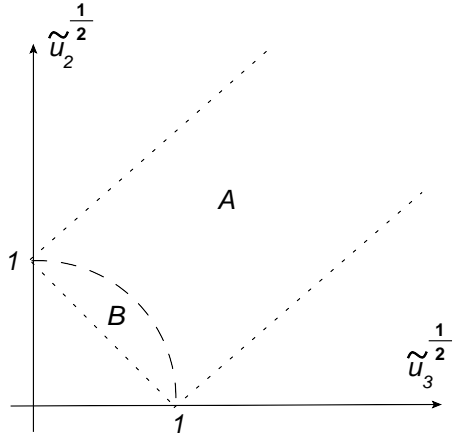,width=60mm}
	\end{center}
	\caption{The region of the reduced cross ratios where the analytic continuation is performed.   }
	\label{fig:region}
\end{figure}

Due to the M\"obius invariance we can put
\beq
x_{A}=1,\;\;\; x_{B}=0,\;\;\; x_{0'}=\infty,\;\;x_{0}=z,
\eeq
then
\beq\label{Kvarcomplex}
\tilde{u}_2=|z|^2, \;\;\; \tilde{u}_3=|1-z|^2,
\eeq
for $z$ given by Eq.~\ref{Kcomplex}.
This imposes a restriction on the possible values of the reduced cross ratios as  illustrated in Fig.~\ref{fig:region}  (see ref.~\cite{Lipatov:2010qg} for more details).

The region $\mathbf{A}$ in  Fig.~\ref{fig:region}  is the region of possible values of the reduced cross ratios, which correspond to the particle  momentum parametrization. Its subregion $\mathbf{B}$ is the region, where the analytic continuation is performed. As it was already mentioned we made sure that our result is valid in the whole region $\mathbf{A}$.

\begin{figure}[htbp]
	\begin{center}
		\epsfig{figure=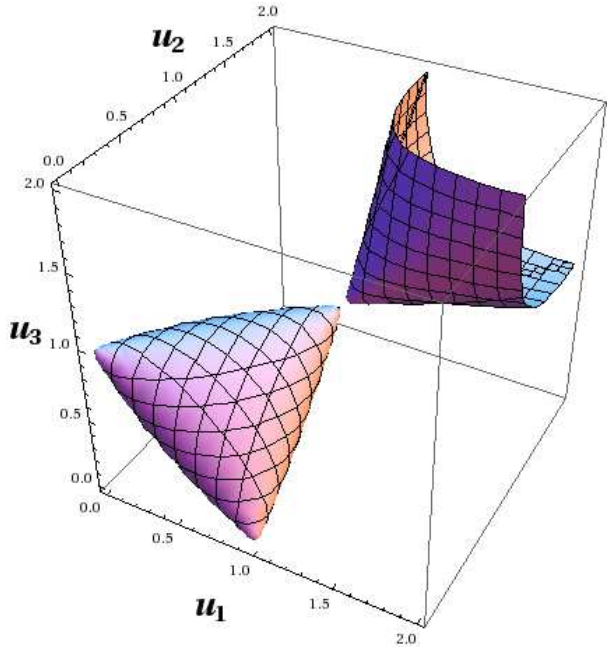,width=80mm}
	\end{center}
	\caption{ The surface $\Delta =0$ is the boundary  of the physical values of  cross ratios.    }
	\label{fig:bubble}
\end{figure}

The same conclusion concerning the physical region of the cross ratios can be reached using the parametrization of the cross ratios introduced by  Goncharov,  Spradlin,  Vergu and  Volovich~\cite{Goncharov:2010jf}. One can parametrize the cross ratios by six complex variables, namely
\beqn\label{zparam}
u_1=\frac{z_{23}z_{56}}{z_{25}z_{36}},\;\;\;
 u_2=\frac{z_{16}z_{34}}{z_{14}z_{36}}, \;\;\;
  u_3=\frac{z_{12}z_{45}}{z_{14}z_{25}},
\eeqn
where $z_{ij}=z_i-z_j$. In this parametrization the square roots in the arguments of the remainder function Eq.~\ref{R6} disappear and $x^{\pm}_{i}$ are rational functions.  Exploiting the conformal invariance we can set
\beqn
z_4=0, \;\;\; z_5=1, \;\;\; z_6=\infty.
\eeqn
Then Eq.~\ref{zparam} reads
\beqn\label{zparam2}
u_1=\frac{z_3-z_2}{1-z_2},\;\;\;
 u_2=\frac{z_3}{z_2}, \;\;\;
  u_3=\frac{z_1-z_2}{z_1(1-z_2)}.
\eeqn
Solving Eq.~\ref{zparam2} for $z_i$ we obtain some square roots that determine the physical region. For example, one of the solutions is given by
\beqn
z_1=\frac{-1+u_1+u_2+u_3 \pm \sqrt{(1-u_1-u_2-u_3)^2-4u_1u_2u_3}}{2 u_2 u_3}
\eeqn
The argument of the square root coincides with $\Delta=(1-u_1-u_2-u_3)^2-4u_1u_2u_3$ defined in Eq.~\ref{Xpm}.
The surface $\Delta=0$ determines the boundary of the space of the physical values of the cross ratios. This surface is depicted in Fig.~\ref{fig:bubble}.

\begin{figure}[htbp]
	\begin{center}
	\subfigure[]{
	\epsfig{figure=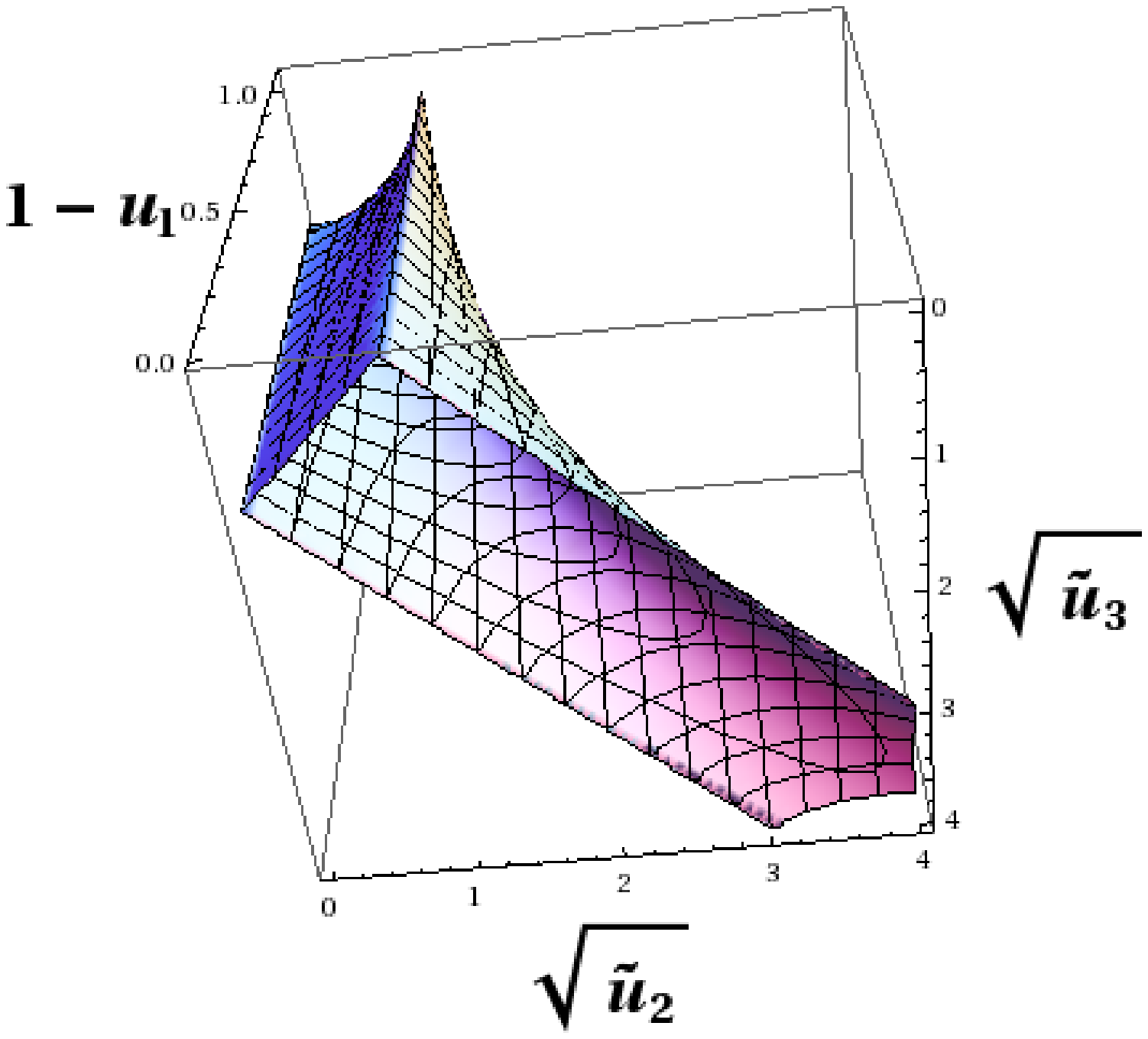,width=70mm}}
	\subfigure[]{
	\epsfig{figure=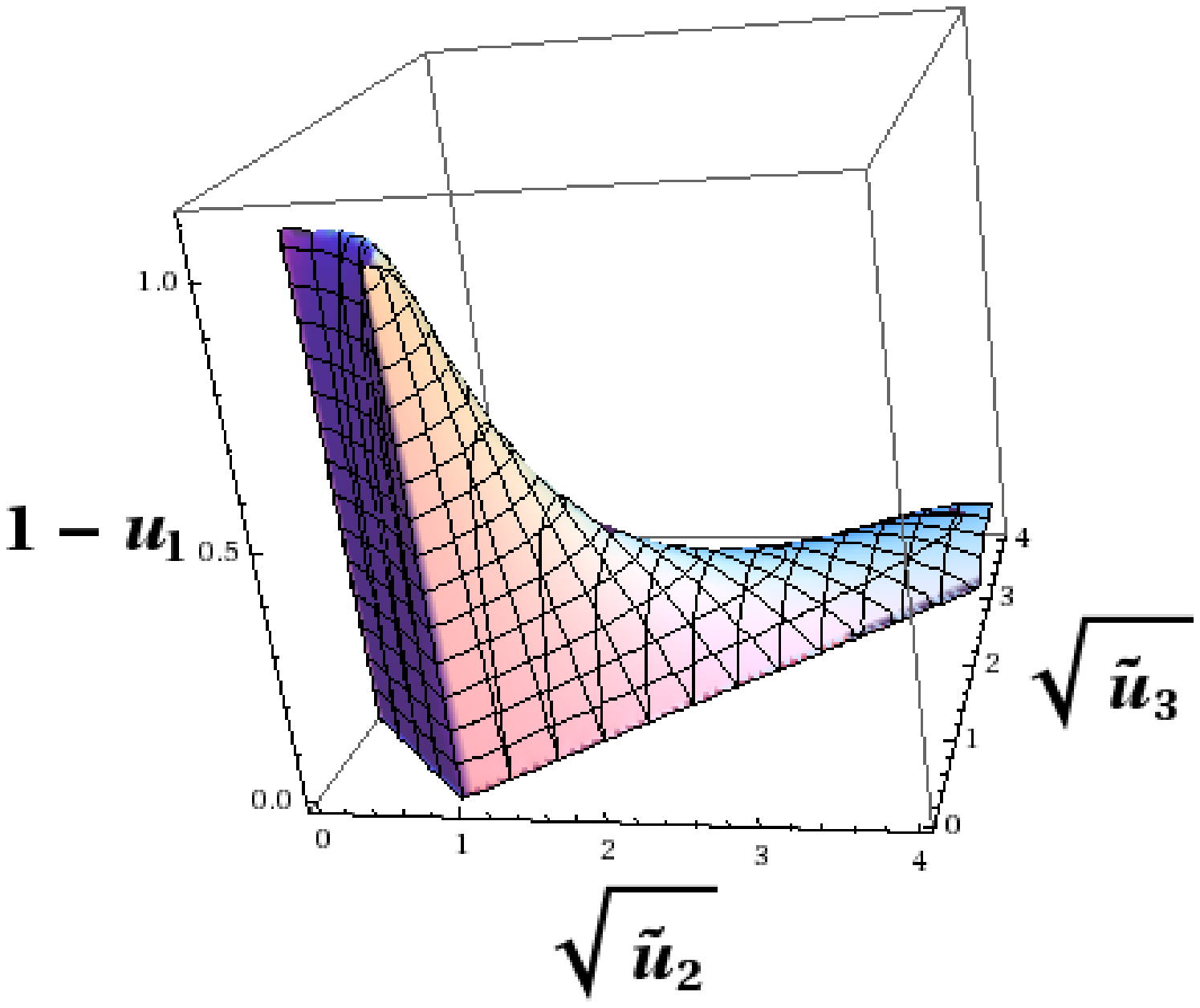,width=70mm}}
		\end{center}
	\caption{ The surface $\Delta=0$ in the coordinates $1-u_1$, $\sqrt{\tilde{u}_2}$ and  $\sqrt{\tilde{u}_3}$. In the Regge limit, when $(\mathbf{k}_1+\mathbf{k}_2)^2/s_2\simeq 1-u_1 \to 0$ we obtain the region $\mathbf{A}$ in Fig.~\ref{fig:region}. Once the Regge kinematics is relaxed we get a physical region of a  cigar shape  instead of the semi-infinite strip in Fig.~\ref{fig:region}. The figures $(\mathbf{a})$ and $(\mathbf{b})$ depicts a different view of the same surface $\Delta=0$. }
	\label{fig:bubbletilde}
\end{figure}

The same plot, but for $AdS_4$ surface was obtained by Alday, Gaiotto and Maldacena~\cite{Alday:2009dv} introducing momenta parametrization of the cross ratios. The space of the physical cross ratios in the unit cube is inside the "bag" in Fig.~\ref{fig:bubble}. To find a match between our picture of the physical region depicted  in Fig.~\ref{fig:region}   and the  "bag" in Fig.~\ref{fig:bubble} we   draw the surface $\Delta=0$ in the coordinates $1-u_1$, $\sqrt{\tilde{u}_2}$ and  $\sqrt{\tilde{u}_3}$ as illustrated in Fig.~\ref{fig:bubbletilde}.

It is clear from Fig.~\ref{fig:bubbletilde}a that in the Regge limit, when $(\mathbf{k}_1+\mathbf{k}_2)^2/s_2\simeq 1-u_1 \to 0$ the boundary of the surface becomes a semi-infinite strip in accordance with Fig.~\ref{fig:region}. If one relaxes the Regge kinematics this region becomes a closed cigar shaped region as follows from the geodesics  in Fig.~\ref{fig:bubbletilde}a.

In the next section we calculate the next-to-leading-order impact factor from our result of the analytic  continuation of the remainder function in Eq.~\ref{KR62result}.

\section{NLO impact factor}\label{sec:NLO}

In this section we calculate the  next-to-leading~(NLO) impact factor appearing in the BFKL approach. We begin with Eq.~\ref{corLLA} and Eq.~\ref{LLA}, which define the imaginary part of the  BDS violating term with logarithmic accuracy. We are interested in  generalizing Eq.~\ref{LLA} to include also next-to-leading in logarithm of the energy~(NLLA) terms. Taking into account NLLA corrections corresponds to relaxing multi-Regge kinematics to quasi-multi-Regge kinematics~(QMRK) for the intermediate particles in the unitarity relation for the amplitude.

 The next-to-leading corrections to Eq.~\ref{LLA} are of two kinds: the  NLO corrections to the impact factors of the BFKL ladder
 and to the kernel of the BFKL equation. The later was calculated in QCD by Fadin and Fiore~\cite{Fadin:2004zq,Fadin:2005zj} and in  $\mathcal{N}=4$ SYM  it  can be extracted from QCD calculations  applying a principle of the maximal transcendentality proposed by Kotikov and one of the authors~\cite{Kotikov:2002ab}. The maximal transcendentality principle was successfully used, for example, to predict   the  anomalous dimension up to six loops  in  $\mathcal{N}=4$ SYM~\cite{Kotikov:2001sc,Kotikov:2003fb,Kotikov:2004er,Kotikov:2006ts,Kotikov:2007cy,Lukowski:2009ce,Velizhanin:2010cm}. Another sort of NLO corrections to Eq.~\ref{LLA} is the corrections to the Reggeon-Reggeon-Particles~(RRP) impact factor of the BFKL ladder, which were never calculated before for the octet channel\footnote{The NLO corrections to the impact factor in the singlet channel were found by    Balitsky and Chirilli~\cite{Balitsky:2009yp,Balitsky:2010ze}.}. It worth mentioning that there is some ambiguity in the higher order terms, namely, the corrections can be redistributed between the impact factor and the BFKL Kernel, which does not affect the form of the amplitude provided the corrections are included in a consistent way.

We are interested in the NLO contribution to the impact factor. We  write Eq.~\ref{LLA}  for the leading logarithmic contribution as

\beqn\label{LLA2}
&& \Delta _{2\rightarrow 4}=\frac{a}{2}\, \sum _{n=-\infty}^\infty (-1)^n
\int _{-\infty}^\infty \frac{d\nu }{\nu ^2+\frac{n^2}{4}}\,
\left(\frac{q_3^*k^*_1}{k^*_2q_1^*}\right)^{i\nu -\frac{n}{2}}\,
\left(\frac{q_3k_1}{k_2q_1}\right)^{i\nu +\frac{n}{2}}\,
\left(s_2^ {\omega (\nu , n)}-1\right)\, \nonumber \\
&& \hspace{3cm}=
\frac{a}{2}\, \sum _{n=-\infty}^\infty
\int _{-\infty}^\infty d\nu \; (2{\chi_1}^{LLA})(2{\chi_2}^{LLA})\;
\left((1-u_1)^ {-\omega (\nu , n)}-1\right),
\eeqn
where the leading-log impact factors are given by
\beqn\label{chiLLA}
&& \chi^{LLA}_1= \frac{1}{2}\frac{1}{\left(i\nu +\frac{n}{2}\right)}\left(-\frac{q_1}{k_1}\right)^{-i\nu -\frac{n}{2}}
\left(-\frac{q^*_1}{k^*_1}\right)^{-i\nu +\frac{n}{2}}, \nonumber\\
&& \chi^{LLA}_2= -\frac{1}{2}\frac{1}{\left(i\nu -\frac{n}{2}\right)}\left(\frac{q^*_3}{k^*_2}\right)^{i\nu -\frac{n}{2}}
\left(\frac{q_3}{k_2}\right)^{i\nu +\frac{n}{2}}.
\eeqn

The functions $\chi^{LLA}_1$ and $\chi^{LLA}_2$  are a  convolution of the octet BFKL eigenfunction and the corresponding impact factor, but for the purpose of our discussion we call them impact factors in the $\nu, n$ representation. For more details regarding the rigorous definition of the impact factors the reader is referred to ref.~\cite{BLS2}. The factor of two accompanying $\chi^{LLA}_i$ in Eq.~\ref{LLA2} is introduced to match the notation in Eqs.~90-93 of ref.~\cite{BLS2}.

In the appendix~\textbf{F} we found that the NLLA  term   of the remainder function at two loops in Eq.~\ref{R62w} can be written as

\beqn \label{NLORtext}
&& R^{(2)\;NLLA}_6\left(|u_1|e^{-i2\pi},\frac{1}{|1+w|^2},\frac{|w|^2}{|1+w|^2}\right) \\
&&= \sum_{n=-\infty}^{\infty} \int d\nu
\frac{i}{2}\frac{(-1)^n}{\nu^2+\frac{n^2}{4}}\left(E^2_{\nu,n}-\frac{1}{4}\frac{n^2}{ \left(\nu^2+\frac{n^2}{4}\right)^2} \right)
\,
\left(w^*\right)^{i\nu -\frac{n}{2}}\,
\left(w\right)^{i\nu +\frac{n}{2}}\,\;\;\;\nonumber \\
&&= \sum_{n=-\infty}^{\infty} \int d\nu
\frac{i}{2}\frac{(-1)^n}{\nu^2+\frac{n^2}{4}}\left(E^2_{\nu,n}-\frac{1}{4}\frac{n^2}{ \left(\nu^2+\frac{n^2}{4}\right)^2} \right)
\,
\left(\frac{q_3^*k^*_1}{k^*_2q_1^*}\right)^{i\nu -\frac{n}{2}}\,
\left(\frac{q_3k_1}{k_2q_1}\right)^{i\nu +\frac{n}{2}}\,\;\;\;\nonumber.
\eeqn
This allows to modify the leading logarithmic expression Eq.~\ref{LLA2} to account  for the next-to-leading in $\ln s_2$~(NLLA) corrections at all orders of the perturbation theory.

Before we proceed there is one fine point to be clarified. According to the factorization hypothesis~\cite{Alday:2007hr} the all-order remainder function $R$ is defined by
\beqn
M_{2\to 4}=M^{BDS}_{2\to 4} R
\eeqn
It was argued by one of the authors~\cite{Lipatov:2010qf} that provided the factorization hypothesis holds, a significant information about the remainder function can be obtained from the analytic properties of the BDS formula. In particular, in the region under consideration, where $s,s_2>0$ and  $s_1,s_3,s_{012},s_{123}<0$ in the Regge kinematics the remainder function at all orders of the perturbation theory is given by the dispersion relation (see\footnote{The  function $R$ is denoted by $c$, and  the reduced cross ration $\tilde{u}_2$ and $\tilde{u}_3$ are  $\phi_2$ and  $\phi_3$ in the notation of ref.~\cite{Lipatov:2010qf}.} Eq.~50 of ref.~\cite{Lipatov:2010qf})
\beqn\label{disp}
R \,e^{i\pi \,\delta}=\cos \pi \omega _{ab}+i\int _{-i\infty}^{i \infty}\frac{d\omega}{2\pi i}\,
f(\omega )\,e^{-i\pi \omega}\,(1-u_1)^{-\omega}\,,
\eeqn
where the phases $\delta$ and $\omega_{ab}$ represent contribution of Regge poles  obtained directly from the BDS formula and are given by
\beqn\label{deltaomega}
\delta =\frac{\gamma _K}{8}\,\ln (\tilde{u}_2 \tilde{u}_3)\,,\,\,
\omega _{ab}
=\frac{\gamma _K}{8}\,\ln \frac{\tilde{u}_3}{\tilde{u}_2}.\,
\eeqn
The coefficient  $\gamma _K$ is the cusp anomalous dimension known to any order of the perturbation theory. The only unknown piece in Eq.~\ref{disp} is the real function $f(\omega)$, which has the Mandelstam cut in $\omega$ and depends only on the transverse momenta and has no energy dependence. In the leading logarithmic approximation $f(\omega)$ is given by
\beqn\label{fomegaLLA}
&&f^{LLA}(\omega)=\frac{a}{2}\sum_{n=-\infty}^{\infty} \int d \nu \frac{1}{\omega-\omega(\nu,n)} \frac{(-1)^n}{\nu^2+\frac{n^2}{4}}\left(\frac{q_3^*k^*_1}{k^*_2q_1^*}\right)^{i\nu -\frac{n}{2}}\,
\left(\frac{q_3k_1}{k_2q_1}\right)^{i\nu +\frac{n}{2}}\\
&&=\frac{a}{2}\sum_{n=-\infty}^{\infty} \int d \nu \frac{1}{\omega-\omega(\nu,n)} (2\chi_1^{LLA})(2\chi_2^{LLA})\nonumber,
\eeqn
where $\omega(\nu,n)$ is defined in Eq.~\ref{eigen}.
 Therefore in RHS of Eq.~\ref{disp} the integral over $\omega$ gives correctly the leading asymptotics of imaginary and real parts of the amplitudes~(see Eq.~\ref{LLA}).

 The expression for $f^{LLA}(\omega)$ is read out from Eq.~\ref{LLA2} and can be generalized to include subleading contributions. By analogy with Eq.~\ref{fomegaLLA} we write
\beqn\label{fomegaNLLA}
f^{NLLA}(\omega)=\frac{a}{2}\sum_{n=-\infty}^{\infty} \int d \nu \frac{1}{\omega-\tilde{\omega}(\nu,n)} (2\tilde{\chi_1})(2\tilde{\chi_2}),
\eeqn
where both the impact factors $\tilde{\chi}_i$ and the BFKL energy $\tilde{\omega}(\nu,n)$ include the next-to-leading corrections. They are defined by
\beqn
\tilde{\chi}_i=\chi^{LLA}_i+\chi^{NLO}_i
\eeqn
and
\beqn
\tilde{\omega}(\nu,n)=\omega(\nu,n)+\omega^{NLO}(\nu,n).
\eeqn

The expression for $\omega^{NLO}(\nu,n)$ can be found from the next-to-leading corrections to the octet BFKL Kernel calculated by Fadin and Fiore~\cite{Fadin:2004zq,Fadin:2005zj}, and the missing NLO impact factors we can readily extract from Eq.~\ref{NLORtext}. In the appendix~\textbf{F} we find that the next-to-leading impact factors $\chi^{NLO}_i$ are given by
\beqn \label{NLOchi1}
\chi^{NLO}_1= \frac{a}{4}\frac{1}{\left(i\nu +\frac{n}{2}\right)}
\left(E^2_{\nu,n}-\frac{1}{4}\frac{n^2}{ \left(\nu^2+\frac{n^2}{4}\right)^2} \right)
\left(-\frac{q_1}{k_1}\right)^{-i\nu -\frac{n}{2}}
\left(-\frac{q^*_1}{k^*_1}\right)^{-i\nu +\frac{n}{2}}
\eeqn
and
\beqn\label{NLOchi2}
\chi^{NLO}_2= -\frac{a}{4}\frac{1}{\left(i\nu -\frac{n}{2}\right)}
\left(E^2_{\nu,n}-\frac{1}{4}\frac{n^2}{ \left(\nu^2+\frac{n^2}{4}\right)^2} \right)
\left(\frac{q_3}{k_2}\right)^{i\nu -\frac{n}{2}}
\left(\frac{q^*_3}{k^*_2}\right)^{i\nu +\frac{n}{2}},
\eeqn
where $E_{\nu,n}$ is defined in Eq.~\ref{Enun}.
 An important feature of $\chi^{NLO}_i$ is that both Eq.~\ref{NLOchi1} and  Eq.~\ref{NLOchi2} do not have holomorphic separability, namely  either $i\nu +n/2$ or $-i\nu +n/2$ cannot be assigned only to one of the projectiles. It is worth emphasizing that the NLO impact factors $\chi^{NLO}_i$ are factorized in the product of the Born impact factors in Eq.~\ref{chiLLA} and a term expressed through the eigenvalue $E_{\nu,n}$ of the BFKL equation in LLA. The form of the NLO impact factor in the $\nu,n$ representation resembles the three-loop remainder function in LLA, emphasizing the intimate relation between the two as discussed in the next section.

\section{Three loops in LLA and NLLA}\label{sec:3loops}

 In this section we calculate the three-loop leading logarithmic~(LLA) contribution to the remainder function of the six-point MHV amplitude and find the real part of the subleading in $\ln s_2/s_0$~(NLLA) term. In the leading logarithm approximation~(LLA) one neglects all terms, which are not enhanced by the logarithm of energy. In our case the main contribution comes from the Mandelstam cut in the variable $\omega$ canonically conjugated to  $\ln s_2$~(see Fig.~\ref{fig:6pointI}) and the leading terms are those which have  each power of the coupling constant   accompanied by the same power of the logarithm of the energy $\ln s_2/s_0$ or equivalently by $\ln(1-u_1)$, because in the Regge limit $1-u_1 \simeq (\mathbf{k}_1+\mathbf{k}_2)^2/s_2$. The all-loop LLA contribution to the remainder function in the Mandelstam channel (see Fig.~\ref{fig:6pointI}b) is given by Eq.~\ref{corLLA} and Eq.~\ref{LLA}. The three-loop term is obtained expanding $s_2^{\omega(\nu,n)}$ in Eq.~\ref{LLA} in powers of the coupling $a$ and reads
 \beqn \label{3loopE2}
&& R_{6}^{(3)\;LLA}=\frac{i\Delta^{(3)}_{2\to 4}}{a^3}=\frac{i}{4}\ln^2(1-u_1)\sum_{n=-\infty}^{\infty} \int d \nu \frac{(-1)^n}{\nu^2+\frac{n^2}{4}}E^2_{\nu,n}
\left(\frac{q^*_3k^*_1}{k^*_2q^*_1}\right)^{i\nu-\frac{n}{2}}
 \left(\frac{q_3k_1}{k_2q_1}\right)^{i\nu+\frac{n}{2}} \\
 && =\frac{i}{4}\ln^2 (1-u_1)\sum_{n=-\infty}^{\infty} \int d \nu \frac{(-1)^n}{\nu^2+\frac{n^2}{4}}E^2_{\nu,n}
\left(w^*\right)^{i\nu-\frac{n}{2}}
 \left(w \right)^{i\nu+\frac{n}{2}}, \nonumber
 \eeqn
where the complex variable $w$ is defined by
\beqn
w=\frac{q_3k_1}{k_2q_1},\;\;\; w^*=\frac{q^*_3k^*_1}{k^*_2q^*_1}.
\eeqn
The expression in Eq.~\ref{3loopE2} is calculated in the appendix~\ref{app:3loops} and has the form
\beqn\label{R63appText}
&&  R_{6}^{(3)\;LLA}=i\Delta^{(3)} _{2\rightarrow 4}/a^3=i\pi \frac{1}{4} \ln^2(1-u_1)\left(
 \ln|w|^2\ln^2|1+w|^2-\frac{2}{3}\ln^3|1+w|^2\right. \hspace{1cm}\;\;\;
\\
&&\left.-\frac{1}{4}\ln^2|w|^2 \ln|1+w|^2+\frac{1}{2} \ln|w|^2 \left(\text{Li}_2(-w)+\text{Li}_2(-w^*)\right)
 - \text{Li}_3(-w)-\text{Li}_3(-w^*)\right). \nonumber
\eeqn

The LLA three-loop remainder function in Eq.~\ref{R63appText} is pure imaginary and vanishes at $w \to 0$ and $w \to \infty$. It is invariant under $w \to 1/w$ transformation, which corresponds to the target-projectile symmetry.

The next-to-leading in the logarithm of the energy $\ln s_2/s_0\simeq -\ln(1-u_1)$ contribution can be obtained from Eq.~\ref{disp} and Eq.~\ref{fomegaNLLA}.  Unfortunately we do not have an explicit expression for $\omega(\nu,n)$ beyond the leading order, which is necessary for this calculation. However it turns out that we can find  asymptotic behavior of the real part of the NLLA remainder function at three loops, because it does not require any knowledge of the higher order corrections to $\omega(\nu,n)$.  $\Re\left(R^{(3)\;NLLA}_6\right)$is calculated expanding Eq.~\ref{disp} in powers of $a$. The details of this analysis are presented in the appendix~\ref{app:real3NLLA} and the result is given by
\beqn\label{R63NLLAtext}
&&  \Re(R^{(3)\;NLLA}_6)= \frac{\pi^2}{4} \ln (1-u_1)\left(
 \ln|w|^2\ln^2|1+w|^2-\frac{2}{3}\ln^3|1+w|^2\right. \hspace{1cm}\;\;\;
\\
&&\left.-\frac{1}{2}\ln^2|w|^2 \ln|1+w|^2- \ln|w|^2 \left(\text{Li}_2(-w)+\text{Li}_2(-w^*)\right)
 +2 \text{Li}_3(-w)+2\text{Li}_3(-w^*)\right). \nonumber
 \eeqn

Note that $\Re(R^{(3)\;NLLA}_6)$ resembles very much the form of $R^{(3)\;LLA}$ in Eq.~\ref{R63app} as one could expect from Eq.~\ref{reNLLAeq}.
The complex variables $w$  is expressed in terms of the reduced cross ratios  $\tilde{u}_2=u_2/(1-u_1)$ and $\tilde{u}_3=u_3/(1-u_1)$ as
 \beqn
w=\frac{1-z}{z}=\frac{B^+}{\tilde{u}_2},\;\;\; w^*=\frac{1-z^*}{z^*}=\frac{B^-}{\tilde{u}_2}
 \eeqn
for $B^{\pm}$ defined in Eq.~\ref{Bpm} by
\begin{eqnarray}
B^{\pm}=\frac{1-\tilde{u}_2-\tilde{u}_3\pm \sqrt{(1-\tilde{u}_2-\tilde{u}_3)^2-4\tilde{u}_2\tilde{u}_3}}{2}.
\end{eqnarray}

\section{Conclusion}
In this paper we present some details of the analytic continuation of the GSVV~\cite{Goncharov:2010jf} formula for the two-loop remainder function $R_{6}^{(2)}$ to a physical region of the $2 \to 4$ particle scattering, where $R_{6}^{(2)}$ gives a non-vanishing contribution in the Regge limit.  We find that after the analytic continuation the remainder function reproduces the leading logarithmic~(LLA)  term calculated by Bartels, Sabio Vera and one of the authors~\cite{BLS2} in the BFKL approach. We also find a term subleading in the logarithm of the energy~(NLLA) and  extract   the next-to-leading~(NLO) impact factor used in the BFKL technique.
The BFKL approach allows to calculate the LLA contribution to the remainder function at any order of the perturbation theory. The three loop LLA remainder function as well as the real part of the three-loop NLLA remainder function are calculated in the BFKL technique and  presented in this study.   We find that the all-loop LLA and the two-loop NLLA terms of the remainder function are pure imaginary in the Regge limit, while starting at three loops the NLLA remainder function develops a non-vanishing real part in the Regge limit after the analytic continuation to the relevant physical region.

\section{Acknowledgments}
We deeply indebted to J.~Bartels for the  enlightening remarks and continued support. We thank  J.~Kotanski, A.~Sever and M.~Spradlin for valuable and stimulating comments. A.P. is grateful to G.~P.~Korchemsky for fruitful discussions and hospitality in Saclay, where part of this work has been done. This work was supported by the Russian grant RFBR-10-02-01338-a.

\newpage


\appendix

\setcounter{equation}{0}

\renewcommand{\theequation}{A.\arabic{equation}}
\section{Polylogarithms of $u_i$}
We perform an  analytic continuation  of the GSVV expression Eq.~\ref{R6} in the variable $u_1$.
The functions which do not depend on $u_1$ remain the same after the continuation.  In the multi Regge kinematics given by Eq.~\ref{MRK}, they can be simplified as follows
\begin{eqnarray}\label{Li2oneminusoveru2}
 \text{Li}_2\left(1-\frac{1}{u_2}\right)\simeq-\zeta_2-\frac{1}{2}\ln^2(1-u_1)-\ln(1-u_1)\ln \tilde{u}_2-\frac{1}{2}\ln^2 \tilde{u}_2
\end{eqnarray}
and
\begin{eqnarray}\label{Li4oneminusoveru2}
 && \text{Li}_4\left(1-\frac{1}{u_2}\right) \simeq  -\frac{7\pi^4}{360}-\frac{\zeta_2}{2}\ln^2(1-u_1)-\frac{1}{24}\ln^4(1-u_1) -\zeta_2 \ln(1-u_1) \ln \tilde{u}_2 \\
&& -\frac{1}{6}\ln^3(1-u_1) \ln \tilde{u}_2 -\frac{\zeta_2}{2}\ln^2 \tilde{u}_2-\frac{1}{4}\ln^2(1-u_1)\ln^2 \tilde{u}_2-\frac{1}{6}\ln(1-u_1)\ln^3 \tilde{u}_2-\frac{1}{24}\ln^4 \tilde{u}_2. \nonumber
\end{eqnarray}
in terms of the  reduced cross ratios defined in Eq.~\ref{redcross}.

The expressions for $\text{Li}_2\left(1-\frac{1}{u_3}\right)$ and $ \text{Li}_4\left(1-\frac{1}{u_3}\right)$ are obtained from Eq.~\ref{Li2oneminusoveru2} and Eq.~\ref{Li4oneminusoveru2} replacing $\tilde{u}_2$
by  $\tilde{u}_3$ in the argument.

The polylogarithms $\text{Li}_n\left(1-\frac{1}{u_1}\right)$ should be  analytically continued and in the multi  Regge kinematics are given by
\begin{eqnarray}\label{Li2oneminusoveru1}
 && \text{Li}_2\left(1-\frac{1}{u_1}\right)=-\int_0^{1-\frac{1}{u_1}} \frac{dt}{t}\ln(1-t) \simeq -i2\pi \int_1^{1-\frac{1}{|u_1|}}\frac{dt}{t}=-i2\pi \ln\left(1-\frac{1}{|u_1|}\right) \\
&&=-i2\pi (\ln(1-u_1)+i\pi )= -i2\pi\ln(1-u_1)+2\pi^2.
\nonumber
\end{eqnarray}

Note that we assign a phase $\ln(-1)=+i\pi$ since the argument $1-\frac{1}{u_1}$ goes counterclockwise around the origin as we continue from $u_1=e^{i0}|u_1|$ to  $u_1=e^{-i2\pi}|u_1|$ through $u_1=e^{-i\pi}|u_1|=-|u_1|$. In a similar way we find

\begin{eqnarray}\label{Li4oneminusoveru1}
 && \text{Li}_4\left(1-\frac{1}{u_1}\right)=-\int_0^{1-\frac{1}{u_1}} \frac{dt}{t}\int_0^{t} \frac{dt'}{t'}\int_0^{t'} \frac{dt''}{t''}\ln(1-t'')  \simeq -i2\pi \frac{1}{6}\ln^3\left(1-\frac{1}{|u_1|}\right) \\
&&=-\frac{\pi^4}{3}+i\pi^3 \ln(1-u_1)+\pi^2\ln^2(1-u_1)-\frac{i\pi}{3}\ln^3 (1-u_1).
\nonumber
\end{eqnarray}


\section{Polylogarithms of $x^\pm$ and  $x_i^\pm$}
\setcounter{equation}{0}

\renewcommand{\theequation}{B.\arabic{equation}}

The variables $x^\pm$ are defined as follows
\begin{eqnarray}\label{xipm}
 x^\pm =\frac{u_1+u_2+u_3-1\pm \sqrt{\Delta}}{2u_1u_2u_3}, \;\;\;\;  x_i^\pm=u_ix^\pm,\end{eqnarray}
where $\Delta=(u_1+u_2+u_3-1)^2-4u_1u_2u_3$.

First we consider the logarithm of $x^-/x^+$. This variable goes from the second quadrant of the complex plane (negative real and positive imaginary part) in the counterclockwise direction during the continuation crossing the real axis for the large negative values of the argument, then it again crosses the real axis for small negative values of the variable. Thus  the argument crosses the branch cut twice and $\ln x^-/x^+ $ remains on the same Riemann sheet and does not acquire any imaginary part after the continuation. Namely,
\begin{eqnarray}\label{xpoverxm}
\ln \frac{x^-}{x^+}=\ln \frac{|x^-|}{|x^+|}\simeq \frac{1-\tilde{u}_2-\tilde{u}_3+\sqrt{(1-\tilde{u}_2-\tilde{u}_3)^2-4\tilde{u}_2\tilde{u}_3}}{1-\tilde{u}_2-\tilde{u}_3-\sqrt{(1-\tilde{u}_2-\tilde{u}_3)^2-4\tilde{u}_2\tilde{u}_3}},
\end{eqnarray}
where the notation $|x^\pm|$ denotes the fact that $u_1$ in $x^\pm$ is replaced by $|u_1|$.

For our purposes it is useful to note that $\ln(1-z)$ has the same  cut structure as $\text{Li}_n(z)$.
This can be seen from the series representation
\begin{eqnarray}
 \text{Li}_n(z) \equiv \sum^{\infty}_{k=1}\frac{z^k}{k^n}
\end{eqnarray}
and thus
\begin{eqnarray}
 \text{Li}_1(z) \equiv \sum^{\infty}_{k=1}\frac{z^k}{k}=-\ln(1-z).
\end{eqnarray}

\subsection{Logarithms of  $x_i^{+}x_i^{-}$}

The variables $x_i^{\pm}$ are defined in Eq.~\ref{xipm}. The functions of these variable that are present in the result  of Goncharov et al. are $\text{Li}_n(x_i^{\pm})$ and $\text{Li}_n(1/x_i^{\pm})$ ($n=1...4$).

The logarithm of the product $x_i^+x_i^-$ can be easily analytically continued noting that
\begin{eqnarray}
x_i^+x_i^-=\frac{u_i}{u_{i+1}u_{i+2}}, \;\;\; i=1...3.
\end{eqnarray}
We  readily calculate
\begin{eqnarray}
\ln x_1^+x_1^-=-i2\pi -\ln \tilde{u}_2-\ln \tilde{u}_3-2\ln(1-|u_1|),
\end{eqnarray}
\begin{eqnarray}
\ln x_2^+x_2^-=i2\pi +\ln \tilde{u}_2-\ln \tilde{u}_3,
\end{eqnarray}
\begin{eqnarray}
\ln x_3^+x_3^-=i2\pi -\ln \tilde{u}_2+\ln \tilde{u}_3,
\end{eqnarray}
because $\ln u_1\simeq -i2\pi$ after the continuation in the limit Eq.~\ref{MRK}.

\subsection{Polylogarithms of $x_1^{\pm}$}
We start with $x_1^{\pm}$. During the analytic continuation  $x_1^{+}$ goes from the second quadrant of the complex plane (negative real and positive  imaginary part) in the clockwise direction and then crosses the real axis between $0$ and $1$. Thus $\text{Li}_n(x_1^{+})$ are not changed after continuation and can be simplified as follows.

First we simplify the argument in the  limit Eq.~\ref{MRK} separating the "longitudinal"( $1-u_1$) and the "transverse" ($\tilde{u}_2$ and $\tilde{u}_3$) cross ratios.  We write
\begin{eqnarray}
|x_1^{\pm}|\simeq -\frac{A_1^{\pm}}{1-|u_1|},
\end{eqnarray}
where  $A_1^{\pm}$ is a function of only $\tilde{u}_2$ and $\tilde{u}_3$  and is given by
\begin{eqnarray}\label{A1pm}
A_1^{\pm}=\frac{1-\tilde{u}_2-\tilde{u}_3 \mp \sqrt{(1-\tilde{u}_2-\tilde{u}_3)^2-4\tilde{u}_2\tilde{u}_3}}{2\tilde{u}_2\tilde{u}_3}.
\end{eqnarray}

Using this notation we  can expand the polylogarithms as follows
\begin{eqnarray}
\text{Li}_1(x_1^+)= \text{Li}_1(|x_1^+|)=-\ln(1-|x_1^+|)\simeq -\ln A_1^{+}+\ln(1-u_1),
\end{eqnarray}
\begin{eqnarray}
\text{Li}_2(x_1^+)= \text{Li}_2(|x_1^+|)\simeq -\frac{\pi^2}{6}-\frac{1}{2}\ln^2 A_1^{+}+\ln A_1^{+}\ln(1-u_1)-\frac{1}{2}\ln^2 (1-u_1),
\end{eqnarray}
\begin{eqnarray}
&& \text{Li}_3(x_1^+)= \text{Li}_3(|x_1^+|)\simeq  -\frac{\pi^2}{6}\ln A_1^{+}-\frac{1}{6}\ln^3 A_1^{+}
+\frac{\pi^2}{6}\ln(1-u_1)+\frac{1}{2}\ln^2 A_1^{+}\ln(1-u_1)\;\;\; \\
&&\hspace{2cm} -\frac{1}{2}\ln A_1^{+} \ln^2(1-u_1)+\frac{1}{6}\ln^3 (1-u_1),\nonumber
\end{eqnarray}
\begin{eqnarray}
&& \text{Li}_4(x_1^+)= \text{Li}_4(|x_1^+|)\simeq -\frac{7\pi^4}{360}-\frac{\pi^2}{12}\ln^2 A_1^{+} -\frac{1}{24}\ln^4 A_1^{+} +\frac{\pi^2}{6}\ln A_1^{+} \ln(1-u_1) \;\;\; \\
&&\hspace{2cm} +\frac{1}{6}\ln^3 A_1^{+}\ln(1-u_1)-\frac{\pi^2}{12} \ln^2(1-u_1)-\frac{1}{4}\ln^2 A_1^{+}\ln^2(1-u_1)
  \nonumber\\
&&\hspace{2cm}+\frac{1}{6}\ln A_1^{+} \ln^3 (1-u_1) -\frac{1}{24}\ln^4 (1-u_1). \nonumber
\end{eqnarray}

The variable $x_1^-$ also does not cross the branch cut of the polylogarithm (from $1$ to $+\infty$) since during the continuation it goes from the third quadrant of the complex plane (both the real and the imaginary parts are negative) in the clockwise direction and then crosses the real axis for negative values (never crosses the imaginary axis). The expansion of the polylogarithms of $x_1^{-}$ is obtained from that of $x_1^{+}$ replacing  $A_1^{+}$ by $A_1^{-}$.

The polylogarithms of $1/x_1^{+}$ are analytically continued since the variable goes from the third quadrant of the complex plane in the counterclockwise direction, crosses the imaginary axis and then the real axis behind $1$ when
we go from $u_1=e^{i0}|u_1|$ to $u_1=e^{-i2\pi}|u_1|$.
 The argument crosses the branch cut of $\text{Li}_n(1/x_1^{+})$, and the direction of the rotation determines the sign of the phase of $\ln(1-1/x_1^{+})$ as positive~($+i2\pi$).

Using this notation we  can expand the polylogarithms as follows
\begin{eqnarray}\label{Li11overx1plus}
\text{Li}_1\left(\frac{1}{x_1^+}\right)\simeq -i2\pi,
\end{eqnarray}
\begin{eqnarray}\label{Li21overx1plus}
&& \text{Li}_2\left(\frac{1}{x_1^+}\right)= -\int_0^{\frac{1}{x_1^+}}\frac{dt}{t}\ln(1-t)=-\int_0^{-\frac{1-|u_1|}{A_1^{+}}}\frac{dt}{t}\ln(1-t)
-i2\pi\int_1^{-\frac{1-|u_1|}{A_1^{+}}}\frac{dt}{t} \;\;\; \\
&& \hspace{2cm} \simeq -i2\pi( \ln(1-u_1)-\ln A_1^{+}+i\pi)= -i2\pi \ln(1-u_1)+i2\pi\ln A_1^{+}+2\pi^2. \nonumber
\end{eqnarray}
In the second line of Eq.~\ref{Li21overx1plus} we used the fact that $\ln(-1)=+i\pi$ because the argument $1/x_1^+$ rotates in the  counterclockwise direction around the origin. In a similar way we continue the rest of the polylogarithms. In general we can write

\begin{eqnarray}
&& \text{Li}_n\left(\frac{1}{x_1^+}\right)\simeq -i2\pi\frac{1}{(n-1)!}\left( \ln(1-u_1)-\ln A_1^{+}+i\pi\right)^{n-1}.
\end{eqnarray}

The polylogarithms of $1/x_1^{-}$ are not continued because the argument goes in the counterclockwise direction in the left complex semi plane and never crosses the  imaginary axis and thus also the branch cut of the polylogarithms. In our limit $1/x_1^{-}\to 0$ and thus
\begin{eqnarray}\label{Lin1overx1minus}
 \text{Li}_n\left(\frac{1}{x_1^-}\right)\simeq 0.
\end{eqnarray}

From Eq.~\ref{Li11overx1plus} and Eq.~\ref{Lin1overx1minus} we   calculate
\begin{eqnarray}
 \text{Li}_1\left(\frac{1}{x_1^+}\right)+\text{Li}_1\left(\frac{1}{x_1^-}\right)\simeq -i2\pi,
\end{eqnarray}
which can be checked by the direct calculation eliminating the square roots before the analytic continuation
\begin{eqnarray}
 \text{Li}_1\left(\frac{1}{x_1^+}\right)+\text{Li}_1\left(\frac{1}{x_1^-}\right)=-\ln \left(1-\frac{1}{x_1^+}\right)-\ln\left(1-\frac{1}{x_1^-}\right)=-\ln \frac{(1-u_2)(1-u_3)}{u_1}\simeq -i2\pi.\;\;\;
\end{eqnarray}
This shows that we fix correctly the phase of the argument according to the direction of its rotation during the analytic continuation despite the presence of the square roots.




\subsection{Polylogarithms of $x_2^{\pm}$}
The variable $x_2^{+}$ rotates from the second quadrant of the complex plane (negative real and positive imaginary parts) in the clockwise direction during the continuation,  crossing the real axis in the left semi plane and never crossing the imaginary axis. This means that the polylogarithms of  $x_2^{+}$ remain the same after the analytic continuation since its argument never crosses the branch cut of the polylogarithm (from real values of $1$ to $+\infty$).

On contrary, $x_2^{-}$ rotates from the third quadrant in the counterclockwise direction crosses the imaginary axis and then the real axis beyond the value of $1$, crossing the branch cut of the polylogarithms. Thus the polylogarithms of this argument are to be analytically  continued.

We start with simplifying $\text{Li}_n(x_2^{+})$. For the sake of convenience, similarly to the previous discussion we write it as
\begin{eqnarray}\label{x2pm}
 |x_2^\pm|=A_2^{\pm}=-A_1^{\pm}\tilde{u}_2,
\end{eqnarray}

where
\begin{eqnarray}
A_2^{\pm}=\frac{\tilde{u}_2+\tilde{u}_3-1 \pm \sqrt{(1-\tilde{u}_2-\tilde{u}_3)^2-4\tilde{u}_2\tilde{u}_3}}{2\tilde{u}_3},
\end{eqnarray}
and $A_1^{\pm}$ was defined in Eq.~\ref{A1pm}. We use redundant definitions $A_n^{\pm}$ , which are expressible through each other solely for the sake of transparency of the intermediate calculations. The final result will be simplified using the relations between them.

As we have already mentioned the polylogarithms of $x_2^{+}$ remain the same after the continuation and are given by
\begin{eqnarray}\label{Linx2plus}
 \text{Li}_n\left(x_2^{+}\right)=\text{Li}_n\left(|x_2^{+}|\right)\simeq \text{Li}_n\left(A_2^{+}\right).
\end{eqnarray}

The polylogarithms of $x_2^{-}$  are to be  continued as follows
\begin{eqnarray}\label{Li1x2minus}
 \text{Li}_1\left(x_2^{-}\right)=-\ln(1-x_2^{-})=-\ln(1-A_2^{-})-i2\pi
\end{eqnarray}
and

\begin{eqnarray}
 \text{Li}_2\left(x_2^{-}\right)=-\int_0^{x_2^{-}}\frac{dt}{t}\ln(1-t)\simeq \text{Li}_2\left(A_2^{-}\right)-i2\pi\ln A_2^{-}.
\end{eqnarray}
In general
\begin{eqnarray}
 \text{Li}_n\left(x_2^{-}\right)\simeq \text{Li}_n\left(A_2^{-}\right)-i2\pi\frac{1}{(n-1)!}\ln^{n-1} A_2^{-}.
\end{eqnarray}

Using Eq.~\ref{Linx2plus} and Eq.~\ref{Li1x2minus} we write

\begin{eqnarray}
 \text{Li}_1\left(x_2^{+}\right)+\text{Li}_1\left(x_2^{-}\right)\simeq-\ln(1-A_2^{+}) -\ln(1-A_2^{-})-i2\pi=-i2\pi +\ln \tilde{u}_3,
\end{eqnarray}

where we used the identity

\begin{eqnarray}
 (1-A_2^{+})(1-A_2^{-})=\frac{1}{\tilde{u}_3}.
\end{eqnarray}

On the other hand we can simplify $\text{Li}_1\left(x_2^{+}\right)+\text{Li}_1\left(x_2^{-}\right)$ before the continuation and continue it after that
\begin{eqnarray}
 \text{Li}_1\left(x_2^{+}\right)+\text{Li}_1\left(x_2^{-}\right)=-\ln\left(1-x_2^{+}\right)-\ln\left(1-x_2^{-}\right)=-\ln \frac{(1-u_1)(1-u_3)}{u_1u_3}\simeq -i2\pi +\ln\tilde{u}_3.\;\;\;
\end{eqnarray}
This confirms our choice of the phase sign according to the direction of the rotation of the argument.

The variable $1/x_2^{+}$ goes from the third quadrant of the complex plane in the clockwise direction during the continuation.
It crosses the real axis in the left complex semi plane and thus never crosses the branch cut of the polylogarithms. The argument
$1/x_2^{-}$ goes from the second quadrant  of the complex plane in the clockwise direction during the continuation. It crosses the imaginary axis and the real axis, but never reaches the branch cut in our limit. Thus  all polylogarithms of $1/x_2^{\pm}$ remain the same after the continuation and can be written as follows
\begin{eqnarray}
 \text{Li}_n\left(\frac{1}{x_2^{\pm}}\right)\simeq \text{Li}_n\left(\frac{1}{A_2^{\pm}}\right).
\end{eqnarray}
As an example we  calculate
\begin{eqnarray}
 \text{Li}_1\left(\frac{1}{x_2^{+}}\right)+\text{Li}_1\left(\frac{1}{x_2^{-}}\right)=-\ln\left(1-\frac{1}{A_2^{+}}\right)-\ln\left(1-\frac{1}{A_2^{-}}\right)\simeq\ln\tilde{u}_2.\;\;\;
\end{eqnarray}

We can check directly the validity of this result by eliminating the square roots before the continuation
\begin{eqnarray}
 \text{Li}_1\left(\frac{1}{x_2^{+}}\right)+\text{Li}_1\left(\frac{1}{x_2^{-}}\right)=-\ln\left(1-\frac{1}{x_2^{+}}\right)-\ln\left(1-\frac{1}{x_2^{-}}\right)=-\ln \frac{(1-u_1)(1-u_3)}{u_2}\simeq \ln \tilde{u}_2.
\end{eqnarray}




\subsection{Polylogarithms of $x_3^{\pm}$}
The polylogarithms of $x_3^{\pm}$ are analytically continued exactly in the same way as corresponding polylogarithms of $x_2^{\pm}$. This can be shown by numerical calculations as well as explained on general grounds by the symmetry $\tilde{u}_2\Leftrightarrow\tilde{u}_3$, which corresponds to target-projectile symmetry of the scattering amplitude.

The polylogarithms of $x_3^{\pm}$ are obtained from that of $x_2^{\pm}$ by making a change $\tilde{u}_2\Leftrightarrow\tilde{u}_3$ as well as replacing $A_2^{\pm}$ by $A_3^{\pm}$, where
\begin{eqnarray}
 A_3^{\pm}=\frac{\tilde{u}_3}{\tilde{u}_2}A_2^{\pm}=-A_1^{\pm}\tilde{u}_3.
\end{eqnarray}




\subsection{Continuation of $J$}
The function $J$ is defined in Eq.~\ref{J} through the sum of the polylogarithm of $x_i^{\pm}$.
Using this definition we can readily find
\begin{eqnarray}
 J=\frac{1}{2}\ln\frac{|x^{-}|}{|x^{+}|}+i\pi\simeq \frac{1}{2}\ln\frac{A_1^{+}}{A_1^{-}}+i\pi.
\end{eqnarray}

As it was anticipated in ref.~\cite{Goncharov:2010jf} this contradicts the result of the continuation of the function
\beq
\ln\frac{x^{-}}{x^{+}}=\ln\frac{|x^{-}|}{|x^{+}|}
\eeq
found in  the previous section (see Eq.~\ref{xpoverxm}) due to the difference in the cut structure between the two. The correct analytic continuation was done using the definition of $J$ in terms of $\text{Li}_1(x_i^{\pm})$ in  Eq.~\ref{J}.




\setcounter{equation}{0}

\renewcommand{\theequation}{C.\arabic{equation}}

\section{Leading Logarithmic Approximation~(LLA)}

We want to extract the leading logarithmic term from the expression of the remainder function after the continuation. The leading logarithm of the energy $\ln s_2$  is related to $\ln(1-u_1)$ through
\begin{eqnarray}
 1-u_1\propto \frac{|\bold{k}_1+\bold{k}_2|^2}{s_2}
\end{eqnarray}
The reduced cross ratios $\tilde{u}_2$ and $\tilde{u}_3$ depend only on the transverse components of the momenta of the external particles and are of the order of unity. It should be emphasized that only terms of the order of $\ln(1-u_1)$ can contribute to the imaginary part of the remainder function at two loops. Higher order terms   $\ln^n(1-u_1)$~($n>1$) would contradict the unitarity of the scattering matrix. These terms do appear at the intermediate steps of the calculations, but they all must cancel out in the final expression.

From our previous discussions we see that only a few terms can have contributions to the  LLA result
\begin{eqnarray}\label{LLAterms}
 -\frac{1}{2}\text{Li}_4\left(1-\frac{1}{u_1}\right)+\text{L}_4(x_1^+,x_1^-)-\frac{1}{8}\left(\sum_{i=1}^3\text{Li}_2\left(1-\frac{1}{u_i}\right)\right)^2.
\end{eqnarray}

We consider them in separate. For our purposes it is convenient to single out only LLA contributions produced in the process of the continuation. We use the fact that the remainder function vanishes   before the continuation
\begin{eqnarray}
R_6^{(2)} \to 0
\end{eqnarray}
 in the limit  $u_1\to 1$, $u_2\to 0$ and $u_3\to 0$ for $\tilde{u}_2=u_2/(1-u_1)$ and $\tilde{u}_3=u_3/(1-u_1)$ being kept fixed and of the order of unity.  We obtain the imaginary part of the remainder function in  LLA by subtracting the contribution before the continuation from  those obtained continuing the functions in the physical region where $u_1 \to e^{-i2\pi}|u_1|$, keeping only terms accompanied by the power of  $\ln(1-u_1)$.
We start with the first term in Eq.~\ref{LLAterms}
\begin{eqnarray}\label{A}
&& -\frac{1}{2}\text{Li}_4\left(1-\frac{1}{u_1}\right)+\frac{1}{2}\text{Li}_4\left(1-\frac{1}{|u_1|}\right)\simeq^{\hspace{-0.35cm}^{LLA}}   -\frac{i\pi^3}{2}\ln(1-u_1)-\frac{\pi^2}{2}\ln^2(1-u_1)+\frac{i\pi}{6}\ln^3(1-u_1),\hspace{0.5cm}
\;\;\;\;
\end{eqnarray}
where we used Eq.~\ref{Li4oneminusoveru1} and the fact that
\begin{eqnarray}\label{Linmodulusu1}
\lim_{u_1\to 1} \text{Li}_n\left(1-\frac{1}{|u_1|}\right)\simeq 0.
\end{eqnarray}

In an analogous way we calculate the second term of  Eq.~\ref{LLAterms}
\begin{eqnarray}\label{B}
 &&\text{L}_4(x_1^+,x_1^-)- \text{L}_4(|x_1^+|,|x_1^-|)\simeq^{\hspace{-0.35cm}^{LLA}}
-\frac{i\pi^3}{3}\ln(1-u_1)-\frac{\pi^2}{2}\ln^2(1-u_1)+\frac{i\pi}{3}\ln^3(1-u_1) \;\;\;\;\;\;\; \\
&& -\frac{\pi^2}{2}\ln(1-u_1)\ln\tilde{u}_2+\frac{i\pi}{2}\ln^2(1-u_1)\ln\tilde{u}_2+\frac{i\pi}{4}\ln(1-u_1)\ln^2\tilde{u}_2-\frac{\pi^2}{2}\ln(1-u_1)\ln\tilde{u}_3\nonumber \\
&&+\frac{i\pi}{2}\ln^2(1-u_1)\ln\tilde{u}_3+\frac{i\pi}{2}\ln(1-u_1)\ln\tilde{u}_2\ln\tilde{u}_3 +\frac{i\pi}{4}\ln(1-u_1)\ln^2\tilde{u}_3.
\nonumber
\end{eqnarray}
The notation $|x_1^\pm|$ denotes that fact that $u_1$ is replaced by $|u_1|$ in the argument.

Finally, the last term in  Eq.~\ref{LLAterms} is given by
\begin{eqnarray}\label{C}
 &&-\frac{1}{8}\left(\sum_{i=1}^3\text{Li}_2\left(1-\frac{1}{u_i}\right)\right)^2-\frac{1}{8}\left(\sum_{i=1}^3\text{Li}_2\left(1-\frac{1}{|u_i|}\right)\right)^2\simeq^{\hspace{-0.35cm}^{LLA}}
\frac{i5\pi^3}{6}\ln(1-u_1)+\pi^2\ln^2(1-u_1) \;\;\;\;\;\;\; \\
&& -\frac{i\pi}{2}\ln^3(1-u_1)+\frac{\pi^2}{2}\ln(1-u_1)\ln\tilde{u}_2-\frac{i\pi}{2}\ln^2(1-u_1)\ln\tilde{u}_2-\frac{i\pi}{4}\ln(1-u_1)\ln^2\tilde{u}_2\nonumber \\
&&+\frac{\pi^2}{2}\ln(1-u_1)\ln\tilde{u}_3-\frac{i\pi}{2}\ln^2(1-u_1)\ln\tilde{u}_3 -\frac{i\pi}{4}\ln(1-u_1)\ln^2\tilde{u}_3.
\nonumber
\end{eqnarray}
Adding up Eq.~\ref{A}, Eq.~\ref{B} and Eq.~\ref{C} we get
\begin{eqnarray}\label{LLAfinal}
 +\frac{i\pi}{2} \ln(1-u_1)\ln \tilde{u}_2 \ln \tilde{u}_3.
\end{eqnarray}
This expression  coincides with LLA term obtained by one of the authors~\cite{BLS2} in the  BFKL approach.




\setcounter{equation}{0}

\renewcommand{\theequation}{D.\arabic{equation}}

\section{Next-to-leading logarithmic~(NLLA) terms}\label{app:NLLA}
We have extracted the leading order term in the logarithm of the energy $\ln s_2 \simeq -\ln(1-u_1)$ of the imaginary part of the remainder function $R_6^{(2)}$. The term we obtained from Eq.~\ref{R6}  after analytic continuation to the physical region of  $u_1 \to e^{-i2\pi}|u_1|$ reproduces the term calculated by one of the authors in the BFKL formalism. Unfortunately, due  complexity of the calculations, the sub-leading terms in $\ln(1-u_1)$ were not yet calculated in the BFKL approach. The comparison between the two approaches, the BFKL formalism and the Wilson Loop/Scattering Amplitude duality, is not full without matching the NLLA terms. In this section we calculated the NLLA terms from the analytically continued expression of Goncharov et al. given in Eq.~\ref{R6} that can be further confronted with the BFKL result once it is available. We follow the logic of the LLA calculations outlined above,  and extract only the NLLA terms that appeared in the course of the continuation (subtracting the relevant values before they were analytically  continued). As it was already mentioned this is possible to do because the remainder function vanishes in the limit of Eq.~\ref{MRK} before the analytic continuation.

The expression of the remainder function is given in Eq.~\ref{R6}. We calculate all contributions in separate leaving only the NLLA terms, i.e. those that are not accompanied by any power of $\ln(1-u_1)$

\begin{eqnarray}\label{Li4ui}
-\frac{1}{2}\sum_{i=1}^3 \text{Li}_4\left(1-\frac{1}{u_i}\right)+\frac{1}{2}\sum_{i=1}^3 \text{Li}_4\left(1-\frac{1}{|u_i|}\right)\simeq^{\hspace{-0.35cm}^{NLO}} \frac{\pi^4}{6},
\end{eqnarray}

\begin{eqnarray}\label{Li2uisquare}
-\frac{1}{8}\left(\sum_{i=1}^3 \text{Li}_2\left(1-\frac{1}{u_i}\right)\right)^2
+\frac{1}{8}\left(\sum_{i=1}^3 \text{Li}_2\left(1-\frac{1}{|u_i|}\right)\right)^2\simeq^{\hspace{-0.35cm}^{NLO}} -\frac{\pi^4}{3}+\frac{\pi^2}{4}\ln^2 \tilde{u}_2+\frac{\pi^2}{4}\ln^2 \tilde{u}_3,
\end{eqnarray}

\begin{eqnarray}\label{L41NLO}
 &&\text{L}_4(x_1^+,x_1^-)- \text{L}_4(|x_1^+|,|x_1^-|)\simeq^{\hspace{-0.35cm}^{NLO}}
 \frac{\pi^4}{8}+\frac{\pi^2}{4}\ln^2 B^{+}+\frac{i\pi}{6}\ln^3 B^{+}-\frac{i \pi^3}{6}\ln (\tilde{u}_2\tilde{u}_3) \\
 && -\frac{\pi^2}{4}\ln B^{+}\ln (\tilde{u}_2\tilde{u}_3)-\frac{i \pi}{4}\ln^2 B^{+}\ln (\tilde{u}_2\tilde{u}_3)
 -\frac{\pi^2}{16}\ln^2 (\tilde{u}_2\tilde{u}_3)  +\frac{i \pi}{8}\ln B^{+}\ln^2 (\tilde{u}_2\tilde{u}_3) +\frac{i \pi}{48}\ln^3 (\tilde{u}_2\tilde{u}_3),
 \nonumber
\end{eqnarray}
where we introduced
\begin{eqnarray}\label{Bpm}
B^{\pm}=\frac{1-\tilde{u}_2-\tilde{u}_3\pm \sqrt{(1-\tilde{u}_2-\tilde{u}_3)^2-4\tilde{u}_2\tilde{u}_3}}{2}
\end{eqnarray}
and used its  property
\begin{eqnarray}
B^{+}B^{-}=\tilde{u}_2\tilde{u}_3
\end{eqnarray}
to eliminate $B^{-}$.

Before we calculate the contributions from $\text{L}_4(x_2^+,x_2^-)$ and $\text{L}_4(x_3^+,x_3^-)$ we find the function $J$.
\begin{eqnarray}\label{JB}
J \simeq \frac{1}{2}\ln \frac{B^{+}}{B^{-}}+i\pi.
\end{eqnarray}
The expression in Eq.~\ref{JB} depends only on the "transverse" cross ratios $\tilde{u}_2$ and $\tilde{u}_3$ as one can see from Eq.~\ref{Bpm}. The function $\chi$ defined in Eq.~\ref{chi} has the same value $\chi=1$ in all points on the circle $u_1=|u_1|e^{i\phi}$ for $|u_1|+u_2+u_3<1$ and thus does not posses any additional terms in the analytic continuation when the phase $\phi$ changes from $0$ to $-i2\pi$. Now we can readily find the contribution from all terms that include $J$

\begin{eqnarray}\label{contriJ}
&&\frac{J^4}{24}+\chi\frac{\pi^2}{12}J^2 -\left(\frac{J^4}{24}+\chi\frac{\pi^2}{12}J^2 \right)|_{u_1=|u_1|}\simeq
-\frac{\pi^4}{24}-\frac{\pi^2}{4}\ln^2 B^{+}+\frac{i\pi}{6}\ln^3 B^{+} \\
&& \hspace{2cm} +\frac{\pi^2}{4}\ln B^{+} \ln(\tilde{u}_2 \tilde{u}_3)-\frac{i\pi}{4}\ln^2 B^{+} \ln(\tilde{u}_2 \tilde{u}_3)
-\frac{\pi^2}{16}  \ln^2(\tilde{u}_2 \tilde{u}_3)\nonumber \\
&& \hspace{2cm} +\frac{i\pi}{8}\ln B^{+} \ln^2(\tilde{u}_2 \tilde{u}_3)-\frac{i\pi}{48}  \ln^3(\tilde{u}_2 \tilde{u}_3).
\nonumber
\end{eqnarray}

Note that
\begin{eqnarray}\label{zeta2}
\chi \frac{\pi^2}{12}\zeta_2-\left(\chi \frac{\pi^2}{12}\zeta_2\right)|_{u_1=|u_1|}=0.
\end{eqnarray}

Summing up Eq.~\ref{Li4ui}-\ref{L41NLO}, Eq.~\ref{contriJ} and Eq.~\ref{zeta2} we obtain a compact expression
\begin{eqnarray}\label{D12379}
-\frac{\pi^4}{12} +\frac{i\pi}{3}\ln^3 B^{+} +\frac{\pi^2}{8}\ln^2\left(\frac{\tilde{u}_2}{\tilde{u}_3}\right)
-\frac{i\pi^3}{6}\ln(\tilde{u}_2\tilde{u}_3)-\frac{i\pi}{2}\ln^2 B^{+}\ln(\tilde{u}_2\tilde{u}_3)
+\frac{i\pi}{4}\ln B^{+}\ln^2(\tilde{u}_2\tilde{u}_3).
\end{eqnarray}

As a last step in our analysis we calculate the contributions of $\text{L}_4(x_2^+,x_2^-)$ and $\text{L}_4(x_3^+,x_3^-)$. Namely,

\begin{eqnarray}\label{L42NLO}
&&\text{L}_4(x_2^+,x_2^-)- \text{L}_4(|x_2^+|,|x_2^-|)\simeq^{\hspace{-0.35cm}^{NLO}} \frac{\pi^4}{24}-\frac{i\pi}{3}\ln^3 B^{+}-\frac{i\pi^3}{6}\ln^3 ( B^{+}+\tilde{u}_2)+\frac{i\pi^3}{6}\ln \tilde{u}_3 \\
&&+\frac{\pi^2}{2} \ln(B^{+}+\tilde{u}_2)\ln \tilde{u}_3 -\frac{\pi^2}{4}\ln^2 \tilde{u}_3+\frac{i\pi}{2} \ln(B^{+}+\tilde{u}_2)\ln^2 \tilde{u}_3
-\frac{i\pi}{6}\ln^3 \tilde{u}_3+\frac{i\pi^3}{12}\ln (\tilde{u}_2\tilde{u}_3)
\nonumber \\
&&+\frac{i\pi}{2}\ln^2 B^{+} \ln(\tilde{u}_2\tilde{u}_3) -\frac{\pi^2}{4}\ln(B^{+}+\tilde{u}_2) \ln(\tilde{u}_2\tilde{u}_3)-\frac{\pi^2}{4}\ln\tilde{u}_3\ln(\tilde{u}_2\tilde{u}_3)
\nonumber\\
&&-\frac{i\pi}{2}\ln(B^{+}+\tilde{u}_2)\ln\tilde{u}_3\ln (\tilde{u}_2\tilde{u}_3)-\frac{i\pi}{4}\ln^2 \tilde{u}_3 \ln(\tilde{u}_2\tilde{u}_3)+\frac{3\pi^2}{16}\ln^2(\tilde{u}_2\tilde{u}_3)-\frac{i\pi}{4}\ln B^{+}\ln^2 (\tilde{u}_2\tilde{u}_3)
\nonumber \\
&&+\frac{i\pi}{8} \ln(B^{+}+\tilde{u}_2)\ln^2 (\tilde{u}_2\tilde{u}_3)+\frac{i3\pi}{8}\ln \tilde{u}_3\ln^2(\tilde{u}_2\tilde{u}_3)-\frac{i\pi}{16}\ln ^3(\tilde{u}_2\tilde{u}_3)-\frac{i\pi^3}{6}\ln(B^{+}+\tilde{u}_3)
\nonumber\\
&&+\frac{\pi^2}{2}\ln \tilde{u}_3 \ln(B^{+}+\tilde{u}_3)+\frac{i\pi}{2}\ln^2 \tilde{u}_3 \ln(B^{+}+\tilde{u}_3)-\frac{\pi^2}{4}\ln (\tilde{u}_2\tilde{u}_3) \ln(B^{+}+\tilde{u}_3)
\nonumber \\
&&-\frac{i\pi}{2}\ln \tilde{u}_3 \ln (\tilde{u}_2\tilde{u}_3)\ln(B^{+}+\tilde{u}_3)+\frac{i\pi}{8}\ln^2(\tilde{u}_2\tilde{u}_3)\ln(B^{+}+\tilde{u}_3)
+\frac{\pi^2}{2}\text{Li}_2\left(-\frac{B^{+}}{\tilde{u}_2}\right) \nonumber \\
&&  +i\pi \ln \tilde{u}_3 \text{Li}_2\left(-\frac{B^{+}}{\tilde{u}_2}\right)-\frac{i\pi}{2}\ln(\tilde{u}_2\tilde{u}_3)\text{Li}_2\left(-\frac{B^{+}}{\tilde{u}_2}\right)
-\frac{\pi^2}{2}\text{Li}_2\left(-\frac{B^{+}}{\tilde{u}_3}\right)-i\pi \ln \tilde{u}_3 \text{Li}_2\left(-\frac{B^{+}}{\tilde{u}_3}\right) \nonumber \\
&& +\frac{i\pi}{2}\ln(\tilde{u}_2\tilde{u}_3)\text{Li}_2\left(-\frac{B^{+}}{\tilde{u}_3}\right) -i\pi \text{Li}_3\left(-\frac{B^{+}}{\tilde{u}_2}\right)
-i\pi \text{Li}_3\left(-\frac{B^{+}}{\tilde{u}_3}\right).
\nonumber
\end{eqnarray}

The contribution of  $\text{L}_4(x_3^+,x_3^-)$ is readily obtained from Eq.~\ref{L42NLO} by changing variables $\tilde{u}_2 \leftrightarrow\tilde{u}_3$.

Summing up Eq.~\ref{D12379} and  Eq.~\ref{L42NLO} (together with $\tilde{u}_2\leftrightarrow\tilde{u}_3$) we get the  NLLA part of the remainder function after the analytic continuation
\begin{eqnarray}\label{NLOfinal}
&&-\frac{i\pi^3}{3}\ln B^{+}-\frac{i\pi}{3}\ln^3 B^{+}-\frac{i\pi}{3}\ln^3 \tilde{u}_2-i\pi \ln B^{+}\ln \tilde{u}_2\ln \tilde{u}_3-\frac{i\pi}{3}\ln^3 \tilde{u}_3 +\frac{i\pi^3}{6}\ln (\tilde{u}_2\tilde{u}_3) \\
&&+\frac{i\pi}{2}\ln^2 B^{+}\ln(\tilde{u}_2\tilde{u}_3)+\frac{i\pi}{6}\ln^3(\tilde{u}_2\tilde{u}_3)
-i\pi \ln\left(\frac{\tilde{u}_2}{\tilde{u}_3}\right)\text{Li}_2\left(-\frac{B^{+}}{\tilde{u}_2}\right)
+i\pi \ln\left(\frac{\tilde{u}_2}{\tilde{u}_3}\right)\text{Li}_2\left(-\frac{B^{+}}{\tilde{u}_3}\right)
\nonumber \\
&& -i2\pi \text{Li}_3\left(-\frac{B^{+}}{\tilde{u}_2}\right)-i2\pi\text{Li}_3\left(-\frac{B^{+}}{\tilde{u}_3}\right).
\nonumber
\end{eqnarray}

The expression of Eq.~\ref{NLOfinal} is pure imaginary in the limit $\tilde{u}_{2,3}>0$ and $\tilde{u}_{2}+\tilde{u}_{3}<1$ despite the fact that it contains a square root in its arguments through $B^{+}$ defined in Eq.~\ref{Bpm}. It is also symmetrical with respect to the exchange of $\tilde{u}_2$ and $\tilde{u}_3$.

 Adding to Eq.~\ref{NLOfinal} the Leading Order result calculated in the previous section and given by Eq.~\ref{LLAfinal} we obtain the final result
\begin{eqnarray}
&&+\frac{i\pi}{2}\ln(1-u_1)\ln \tilde{u}_2\ln \tilde{u}_3 \\
&&-\frac{i\pi^3}{3}\ln B^{+}-\frac{i\pi}{3}\ln^3 B^{+}-\frac{i\pi}{3}\ln^3 \tilde{u}_2-i\pi \ln B^{+}\ln \tilde{u}_2\ln \tilde{u}_3-\frac{i\pi}{3}\ln^3 \tilde{u}_3 +\frac{i\pi^3}{6}\ln (\tilde{u}_2\tilde{u}_3) \nonumber\\
&&+\frac{i\pi}{2}\ln^2 B^{+}\ln(\tilde{u}_2\tilde{u}_3)+\frac{i\pi}{6}\ln^3(\tilde{u}_2\tilde{u}_3)
-i\pi \ln\left(\frac{\tilde{u}_2}{\tilde{u}_3}\right)\text{Li}_2\left(-\frac{B^{+}}{\tilde{u}_2}\right)
+i\pi \ln\left(\frac{\tilde{u}_2}{\tilde{u}_3}\right)\text{Li}_2\left(-\frac{B^{+}}{\tilde{u}_3}\right)
\nonumber \\
&& -i2\pi \text{Li}_3\left(-\frac{B^{+}}{\tilde{u}_2}\right)-i2\pi\text{Li}_3\left(-\frac{B^{+}}{\tilde{u}_3}\right).
\nonumber
\end{eqnarray}
An important remark is in order. The last expression   was calculated in the region $\tilde{u}_2+\tilde{u}_3<1$ and, in principle, should be analytically continued to any other physical region. The first term $+\frac{i\pi}{2}\ln(1-u_1)\ln \tilde{u}_2\ln \tilde{u}_3$, which corresponds to the Leading Logarithmic Approximation is a smooth function also outside the region $\tilde{u}_2+\tilde{u}_3<1$ since it does not have any singularities on the boundary of the region. This is not obvious for the rest of the  NLLA terms, where individual terms do have branch points on the boundary of $\tilde{u}_2+\tilde{u}_3<1$. However the singularities are canceled in the sum as can be shown introducing back $B^{-}$
\begin{eqnarray}\label{NLOfinalApp}
&&R(|u_1|e^{-i2\pi},\tilde{u}_2(1-u_1),\tilde{u}_3(1-u_1))\simeq +\frac{i\pi}{2}\ln(1-u_1)\ln \tilde{u}_2\ln \tilde{u}_3+\frac{i\pi}{3}\ln^3 \tilde{u}_2 \nonumber \\
&&-\frac{i\pi}{2}\ln^2 \tilde{u}_2  \ln\left(\frac{\tilde{u}_2}{\tilde{u}_3}\right)
  -i\pi\ln\left(\frac{\tilde{u}_2}{\tilde{u}_3}\right) \left(\text{Li}_2\left(-\frac{B^{+}}{\tilde{u}_2}\right)+\text{Li}_2\left(-\frac{B^{-}}{\tilde{u}_2}\right)\right)
 \\
&&-i2\pi  \left(\text{Li}_3\left(-\frac{B^{+}}{\tilde{u}_2}\right)+\text{Li}_3\left(-\frac{B^{-}}{\tilde{u}_2}\right)\right).
\nonumber
\end{eqnarray}
It can be easily seen from the series representation of the polylogarithms  that all square roots in the argument cancel out,  and   the expression in  Eq.~\ref{NLOfinalApp} is also valid for $\tilde{u}_2+\tilde{u}_3\geq1$, but only in the  region $\mathbf{A}$ of the multi Regge kinematics shown in Fig.~\ref{fig:region}.  Eq.~\ref{NLOfinalApp} is the main result of this study.




\setcounter{equation}{0}

\renewcommand{\theequation}{E.\arabic{equation}}

\section{$R^{(2)}_6$ in complex variables}\label{app:complex}
In this section we eliminate the square roots in the arguments of the  remainder function of Eq.~\ref{NLOfinalApp} introducing complex variables
\begin{eqnarray}\label{varcomp}
z=\sqrt{\tilde{u}_2}e^{i\phi_2},\;\;\;1-z=\sqrt{\tilde{u}_3}e^{-i\phi_3}.
\end{eqnarray}
It is useful to calculate $\cos (\phi_2-\phi_3)$ and $\sin (\phi_2-\phi_3)$ from
\begin{eqnarray}
 |z+(1-z)|^2=1=\tilde{u}_2+\tilde{u}_3+2\sqrt{\tilde{u}_2}\sqrt{\tilde{u}_3}\cos (\phi_2-\phi_3).
\end{eqnarray}
We    readily find
\begin{eqnarray}\label{cos}
 \cos (\phi_2-\phi_3)=\frac{1-\tilde{u}_2-\tilde{u}_3}{2\sqrt{\tilde{u}_2}\sqrt{\tilde{u}_3}}
\end{eqnarray}

and

\begin{eqnarray}
 \sin (\phi_2-\phi_3)=\frac{\sqrt{4\tilde{u}_2\tilde{u}_3-(1-\tilde{u}_2-\tilde{u}_3)^2}}{2\sqrt{\tilde{u}_2}\sqrt{\tilde{u}_3}}
\end{eqnarray}
as well as
\begin{eqnarray}\label{sin}
 i\sin (\phi_2-\phi_3)=-\frac{\sqrt{(1-\tilde{u}_2-\tilde{u}_3)^2-4\tilde{u}_2\tilde{u}_3}}{2\sqrt{\tilde{u}_2}\sqrt{\tilde{u}_3}}.
\end{eqnarray}
With the help of Eq.~\ref{cos} an Eq.~\ref{sin}  the function  $B^{\pm}$, defined in Eq.~\ref{Bpm}, can be written as
\begin{eqnarray}
 B^{\pm}=e^{\mp i\phi_2}e^{\mp i \phi_3}|z|\;|1-z|
\end{eqnarray}
and thus
\begin{eqnarray}\label{BBZ}
 \frac{B^{+}}{\tilde{u}_2}=\frac{1-z}{z},\;\;\;\frac{B^{-}}{\tilde{u}_2}=\frac{1-z^*}{z^*},\;\;\;\frac{B^{+}}{\tilde{u}_3}=\frac{z^*}{1-z^*},\;\;\;\frac{B^{-}}{\tilde{u}_3}=\frac{z}{1-z}.
\end{eqnarray}
Using Eq.~\ref{BBZ} and the identities between  $\text{Li}_k$ of different arguments we write the expression in Eq.~\ref{NLOfinalApp} as follows
\begin{eqnarray}\label{R62complexApp}
&&R(|u_1|e^{-i2\pi},|z|^2,|1-z|^2)\simeq \frac{i\pi}{2}\ln(1-u_1)\ln |z|^2 \ln |1-z|^2 \\
&&-i4\pi\zeta_3+\frac{i\pi}{2} \ln |z|^2|1-z|^2 \left(\ln z \ln (1-z)+\ln z^* \ln (1-z^*) -2\zeta_2\right)   \nonumber\\
&& +\frac{i\pi}{2} \ln \frac{|1-z|^2}{|z|^2}\left(\text{Li}_2(z)+\text{Li}_2(z^*)-\text{Li}_2(1-z)-\text{Li}_2(1-z^*)\right)
\nonumber \\
&& +i2\pi \left(\text{Li}_3(z)+\text{Li}_3(z^*)+\text{Li}_3(1-z)+\text{Li}_3(1-z^*)\right). \nonumber
\end{eqnarray}
From  Eq.~\ref{R62complexApp} we see that the square roots present in $B^{\pm}$ disappear and   the remainder function is manifestly pure imaginary in LLA. The target-projectile symmetry $\tilde{u}_2 \leftrightarrow \tilde{u}_3$, which is $z\leftrightarrow 1-z$  symmetry in terms of the variables Eq.~\ref{varcomp} is also obvious in Eq.~\ref{R62complexApp}.

Because of the holomorphic factorization of the impact factor in  Eq.~\ref{LLA}  it is more natural to express the final answer in  complex variables
\begin{eqnarray}
 w=\frac{1-z}{z}, \;\; w^*=\frac{1-z^*}{z^*}.
\end{eqnarray}
Noting that the reduced crossed ratios $\tilde{u}_2$ and  $\tilde{u}_3$ are related to the transverse momenta (see Eq.~\ref{redcrosstrans}) we can write
\begin{eqnarray}
\frac{q_3 k_1}{k_2 q_1}=\frac{\sqrt{\tilde{u}_3}e^{i\phi_3}}{\sqrt{\tilde{u}_2}e^{i\phi_2}}=\frac{1-z}{z}=w
\end{eqnarray}

and thus Eq.~\ref{LLA} reads
\beqn\label{LLAz}
&& \Delta _{2\rightarrow 4}=\frac{a}{2}\, \sum _{n=-\infty}^\infty (-1)^n
\int _{-\infty}^\infty \frac{d\nu }{\nu ^2+\frac{n^2}{4}}\,
\left(\frac{1-z^*}{z^*}\right)^{i\nu -\frac{n}{2}}\,
\left(\frac{1-z}{z}\right)^{i\nu +\frac{n}{2}}\,
\left(s_2^ {\omega (\nu , n)}-1\right)\, \nonumber\\
&&
=\frac{a}{2}\, \sum _{n=-\infty}^\infty (-1)^n
\int _{-\infty}^\infty \frac{d\nu }{\nu ^2+\frac{n^2}{4}}\,
\left(w^*\right)^{i\nu -\frac{n}{2}}\,
\left(w\right)^{i\nu +\frac{n}{2}}\,
\left(s_2^ {\omega (\nu , n)}-1\right).\,
\eeqn

Eq.~\ref{LLAz} is  explicitly symmetric in  $z\leftrightarrow 1-z$   ($w\leftrightarrow 1/w$). We recast Eq.~\ref{R62complexApp}  in terms of the variables $w$ and $w^*$ as follows

\begin{eqnarray}\label{R62wApp}
 &&R\left(|u_1|e^{-i2\pi},\frac{1}{|1+w|^2},\frac{|w|^2}{|1+w|^2}\right)\simeq\frac{i\pi}{2}\ln(1-u_1) \ln |1+w|^2 \ln \left|1+\frac{1}{w}\right|^2 \\
&& +\frac{i\pi}{2} \ln |w|^2 \ln^2|1+w|^2-\frac{i\pi}{3}\ln^3 |1+w|^2+ i\pi \ln |w|^2 \left( \text{Li}_2 (-w) +\text{Li}_2 (-w^*)\right)
\nonumber \\
&&-i2\pi  \left( \text{Li}_3 (-w) +\text{Li}_3 (-w^*)\right).
\nonumber
\end{eqnarray}



\setcounter{equation}{0}

\renewcommand{\theequation}{F.\arabic{equation}}

\section{NLO impact factor}\label{sec:complex}
We wish to calculate inverse Mellin and Fourier transforms of the next-to-leading contribution to the remainder function. The form of the direct transforms in the complex variable $w$ can be read out from the last line of Eq.~\ref{LLAz} and is given by

\beqn
 \tilde{f}(w,w^*)=\sum _{n=-\infty}^\infty
\int _{-\infty}^\infty d\nu
\left(w^*\right)^{i\nu -\frac{n}{2}}\,
\left(w\right)^{i\nu +\frac{n}{2}} f\left(\nu,n\right)=
\sum _{n=-\infty}^\infty
\int _{-\infty}^\infty d\nu
\rho^{2i\nu}e^{i\phi n}
f\left(\nu,n\right),
\eeqn
where   $w=\rho e^{i\phi}$.
The inverse transform thus reads
\beqn
f(\nu,n)=\frac{1}{(2\pi)^2}\int_0^\infty  d\rho^2 \int_0^{2\pi}  d\phi \rho^{-2i\nu-2}
e^{-i\phi n} \tilde{f}(\rho,\phi),
\eeqn
which  can be written as
\beqn\label{stamint}
f(\nu,n)=\frac{2}{(2\pi)^2}\int d^2 \vec{w}  (w)^{-i\nu -\frac{n}{2}-\frac{1}{2}} (w^*)^{-i\nu +\frac{n}{2}-\frac{1}{2}} \tilde{f}(w,w^*).
\eeqn
The integration in Eq.~\ref{stamint} is performed on the two dimensional plane in Cartesian coordinates $w_1$ and $w_2$ defined by $w=w_1+iw_2$.
We start with the logarithmic terms appearing in Eq.~\ref{R62wApp}. The relevant logarithms $\ln^k|1+w|^2$ can be obtained by differentiation of the power function $(|1+w|^2)^a$ and thus it is convenient to consider
\beqn\label{loopw}
&&\frac{1}{(2\pi)^2}\int d^2 w\; |w|^{2b} w^{*\;n}|1+w|^{2 a}\\
&&=
\frac{1}{(2\pi)^2}\frac{\Gamma(a+n+1)}{\Gamma(a+1)}\frac{\partial^n }{\partial z^n}\int d^2 w\; |w|^{2b} |z+w|^{2 a+2n}|_{z=z^*=1}\;\;
\nonumber
\eeqn
We introduce the master integral
\beqn\label{master}
g(a,b;z)=\frac{1}{\pi}\int d^2 x|x|^{2a}|z-x|^{2b},
\eeqn
which corresponds to the one loop  diagram with $z$ being a momentum of the external particles.

This integral is found by using the well-known formula of the momentum integration
\beqn
\int \frac{d^d \mathbf{k}}{(\mathbf{k}^2)^{\lambda_1}((\mathbf{q}-\mathbf{k})^2)^{\lambda_2}}=
\pi^{d/2}\frac{\Gamma(d/2-\lambda_1)\Gamma(d/2-\lambda_2)}{\Gamma(\lambda_1)\Gamma(\lambda_2)\Gamma(d-\lambda_1-\lambda_2)}
\frac{\Gamma(\lambda_1+\lambda_2-d/2)}{(\mathbf{q}^2)^{\lambda_1+\lambda_2-d/2}},
\eeqn
and it reads
\beqn
g(a,b;z)=\frac{\Gamma(1+a)\Gamma(1+b)}{\Gamma(-a)\Gamma(-b)\Gamma(2+a+b)}\frac{\Gamma(-1-a-b)}{|z|^{2(-1-a-b)}}.
\eeqn
Using the identity $\Gamma(x)\Gamma(1-x)=\pi/\sin(\pi x)$ this can be written as
\beqn
g(a,b;z)=\frac{\Gamma^2(1+a)\Gamma^2(1+b)}{\Gamma^2(2+a+b)}\frac{\sin \pi a \sin \pi b}{\pi\sin \pi (a+b)}}|z|^{2(1+a+b).
\eeqn

From Eq.~\ref{loopw} and Eq.~\ref{master} we find a general expression for the inverse transform of the logarithms
\beqn\label{Gapp}
\mathcal{G}_{(k,m)}(\nu,n)=\frac{(-1)^n}{2\pi}\frac{\partial^k }{\partial a^k}\frac{\partial^m }{\partial b^m}\frac{\Gamma^2(1+a)\Gamma(1+b)\Gamma(1+b-n)}{\Gamma(2+b+a)\Gamma(2+a+b-n)}\frac{\sin \pi a \sin \pi b}{\pi\sin \pi (a+b)}|_{a=0,b=-i\nu-1+n/2}
\;\;
\eeqn
Applying Eq.~\ref{Gapp} for $k=1$ and $m=0$ we get (for $n\neq 0$)
\beqn
2 \pi \ln|1+w|^2 \Rightarrow -\frac{(-1)^n}{\nu^2+\frac{n^2}{4}}
\eeqn
in full agreement with the Born term of the six-point amplitude.

For $k=2$ and $m=0$ we get from Eq.~\ref{Gapp}
\beqn\label{ln2}
2\pi\ln^2|1+w|^2 \Rightarrow 2\frac{(-1)^n}{\nu^2+\frac{n^2}{4}}\left(
\psi\left(i\nu +\frac{|n|}{2}\right)
+\psi\left(1-i\nu +\frac{|n|}{2}\right)-2\psi\left(1\right)\right),
\eeqn
where we used the identity $\psi(z)=\psi(1-z)-\pi \cot \pi z$.

Plugging  $k=3$ and $m=0$ in Eq.~\ref{Gapp} we get
\beqn\label{ln3}
&& 2\pi \ln^3|1+w|^2 \Rightarrow -\frac{(-1)^n}{\nu^2+\frac{n^2}{4}}\left(-\pi^2+3\left(
\psi\left(i\nu +\frac{|n|}{2}\right)
+\psi\left(1-i\nu +\frac{|n|}{2}\right)-2\psi\left(1\right)\right)^2 \right. \nonumber \\
&& \left.
+3\left(\psi'\left(i\nu +\frac{|n|}{2}\right)-\psi'\left(1-i\nu +\frac{|n|}{2}\right)+2\psi\left(1\right)\right)\right)
\eeqn
and  for  $k=1$ and $m=1$  one obtains
\beqn\label{lnln}
2\pi\ln|w|^2 \ln|1+w|^2 \Rightarrow (-1)^n\frac{2i\nu}{\left(\nu^2+\frac{n^2}{4}\right)^2}.
\eeqn
Therefore from Eq.~\ref{lnln} and Eq.~\ref{ln2} we get
\beqn
&&2\pi\ln |1+w|^2 \ln\left| 1+\frac{1}{w} \right|^2=2\pi\ln^2 |1+w|^2-2\pi\ln |w|^2\ln |1+w|^2\Rightarrow \\
&&2\frac{(-1)^n}{\nu^2+\frac{n^2}{4}}
\left(-\frac{|n|}{2}\frac{1}{\nu^2+\frac{n^2}{4}}+\psi\left(1+i\nu+\frac{|n|}{2}\right)
+\psi\left(1-i\nu+\frac{|n|}{2}\right)-2\psi\left(1\right)\right)=2\frac{(-1)^n}{\nu^2+\frac{n^2}{4}}E_{\nu,n} \nonumber
\eeqn
in an agreement with the  analysis of ref.~\cite{BLS2}.

In a similar way we calculate
\beqn\label{ln1ln2}
&&2\pi\ln|w|^2 \ln^2|1+w|^2 \Rightarrow\\
&&=-(-1)^n \frac{2}{\nu^2+\frac{n^2}{4}}\left(
\frac{-i2\nu}{\nu^2+\frac{n^2}{4}}\left(2\psi\left(1\right)-\psi\left(i\nu+\frac{|n|}{2}\right)-\psi\left(1-i\nu+\frac{|n|}{2}\right)\right)\right.
\nonumber\\
&&\left.-\psi'\left(1-i\nu+\frac{|n|}{2}\right)+\psi'\left(i\nu+\frac{|n|}{2}\right)
\right)\nonumber.
\eeqn

The rest of the terms in Eq.~\ref{R62wApp} are found by noting that
\beqn \label{li2li3}
&&2\pi\ln|w|^2\left(\text{Li}_2(-w)+\text{Li}_2(-w^*)\right)-4\pi\left(\text{Li}_3(-w)+\text{Li}_3(-w^*)\right)
\Rightarrow  -\frac{(-1)^n}{\left(\nu^2+\frac{n^2}{4}\right)^2}. \;\;
\eeqn
This can be easily checked by calculating residue at $\nu=\pm i|n|/2$ and summing over $n$.
Both of the terms on LHS of Eq.~\ref{li2li3} have poles double poles at $\nu=0$ for $n\neq 0$ (for $n=0$ the remainder function vanishes), which correspond to the infrared divergencies absent in the remainder function. Due to the special coefficients of these terms appearing in Eq.~\ref{R62wApp} the infrared divergency is canceled in the final expression. This fact suggests that the relative coefficients of the individual terms can be fixed demanding the absence of the poles at $\nu=0$ for $n\neq0$, together with $w \to 1/w$ symmetry.

Gathering together the inverse transform of all terms in Eq.~\ref{R62wApp}  we finally  obtain
\beqn \label{NLOR}
R^{NLLA}\left(|u_1|e^{-i2\pi},\frac{1}{|1+w|^2},\frac{|w|^2}{|1+w|^2}\right)\Rightarrow
\frac{i}{2}\frac{(-1)^n}{\nu^2+\frac{n^2}{4}}\left(E^2_{\nu,n}-\frac{1}{4}\frac{n^2}{ \left(\nu^2+\frac{n^2}{4}\right)^2} \right),\;\;\;
\eeqn
where $E_{\nu,n}$ is given by Eq.~\ref{Enun}.
This expression vanishes for $n=0$, which corresponds to absence of the infrared divergencies in the remainder function; it is symmetric in $n\to -n$ and $\nu \to -\nu$, which implied by  the target-projectile symmetry $w \to 1/w$.

From the definition of the impact factors in $\nu,n$ representation given by Eq.~\ref{chiLLA} we read out the form of the next-to-leading-order~(NLO) impact factor
\beqn
\chi^{NLO}_1= \frac{a}{4}\frac{1}{\left(i\nu +\frac{n}{2}\right)}
\left(E^2_{\nu,n}-\frac{1}{4}\frac{n^2}{ \left(\nu^2+\frac{n^2}{4}\right)^2} \right)
\left(-\frac{q_1}{k_1}\right)^{-i\nu -\frac{n}{2}}
\left(-\frac{q^*_1}{k^*_1}\right)^{-i\nu +\frac{n}{2}},
\eeqn

\beqn
\chi^{NLO}_2= -\frac{a}{4}\frac{1}{\left(i\nu -\frac{n}{2}\right)}
\left(E^2_{\nu,n}-\frac{1}{4}\frac{n^2}{ \left(\nu^2+\frac{n^2}{4}\right)^2} \right)
\left(\frac{q_3}{k_2}\right)^{i\nu -\frac{n}{2}}
\left(\frac{q^*_3}{k^*_2}\right)^{i\nu +\frac{n}{2}}.
\eeqn

In the next section we use Eq.~\ref{NLOR} to calculate the three loop leading-log contribution to the six-point amplitude.

\setcounter{equation}{0}

\renewcommand{\theequation}{G.\arabic{equation}}

\section{Three loop contribution in LLA}\label{app:3loops}

The general expression for the leading logarithmic contribution to the imaginary part of the remainder function at any number of loops is given by
\beqn
&& \Delta _{2\rightarrow 4}=\frac{a}{2}\, \sum _{n=-\infty}^\infty (-1)^n
\int _{-\infty}^\infty \frac{d\nu }{\nu ^2+\frac{n^2}{4}}\,
\left(\frac{q_3^*k^*_1}{k^*_2q_1^*}\right)^{i\nu -\frac{n}{2}}\,
\left(\frac{q_3k_1}{k_2q_1}\right)^{i\nu +\frac{n}{2}}\,
\left(s_2^ {\omega (\nu , n)}-1\right)\,  \\
&& \hspace{3cm}=
\frac{a}{2}\, \sum _{n=-\infty}^\infty (-1)^n
\int _{-\infty}^\infty \frac{d\nu }{\nu ^2+\frac{n^2}{4}}\,
\left(w^*\right)^{i\nu -\frac{n}{2}}\,
\left(w\right)^{i\nu +\frac{n}{2}}\,
\left(s_2^ {\omega (\nu , n)}-1\right),\,\nonumber\label{LLAapp}
\eeqn
where the $\omega (\nu , n)$ is related to the eigenvalue of the octet BFKL Hamiltonian $E_{\nu,n}$ by
\beqn
\omega (\nu , n)=-a E_{\nu,n}
\eeqn
with
\beqn
E_{\nu,n}=-\frac{1}{2}\frac{|n|}{\nu^2+\frac{n^2}{4}}+\psi\left(1+i\nu+\frac{|n|}{2}\right)+\psi\left(1-i\nu+\frac{|n|}{2}\right)-2\psi\left(1\right).
\eeqn
The complex   variable $w$ is defined through
\beqn\label{wapp}
w=\frac{q_3k_1}{k_2q_1}\;\; \text{and}\;\; w^*=\frac{q^*_3k^*_1}{k^*_2q^*_1}.
\eeqn
According to the discussion presented in appendix \textbf{E} in the Regge limit   $w$ can be written as
\beqn
w=\sqrt{\frac{\tilde{u}_3}{\tilde{u}_2}}e^{i(\phi_3-\phi_2)}
\eeqn
with
\beqn
\cos(\phi_2-\phi_3)=\frac{1-\tilde{u}_2-\tilde{u}_3}{2\sqrt{\tilde{u}_2\tilde{u}_3}}
\eeqn
and
\beqn
\sin(\phi_2-\phi_3)=\frac{\sqrt{4 \tilde{u}_2\tilde{u}_3-(1-\tilde{u}_2-\tilde{u}_3)^2}}{2\sqrt{\tilde{u}_2\tilde{u}_3}}
\eeqn
for
\beqn
\tilde{u}_2=\frac{u_2}{1-u_1}, \;\; \tilde{u}_3=\frac{u_3}{1-u_1}.
\eeqn

The LLA three-loop contribution to the remainder function $R^{(3)}_{6}$ is obtained from Eq.~\ref{LLAapp} by expanding in powers of the coupling constant $a$ as follows
\beqn\label{delta3}
&& \frac{ a^3 R^{(3)\;LLA}_{6}}{i}=\Delta^{(3)} _{2\rightarrow 4}=
\frac{a^3}{4}\ln^2 s_2\, \sum _{n=-\infty}^\infty (-1)^n
\int _{-\infty}^\infty \frac{d\nu }{\nu ^2+\frac{n^2}{4}}\,
\left(w^*\right)^{i\nu -\frac{n}{2}}\,
\left(w\right)^{i\nu +\frac{n}{2}}\,E^2_{\nu,n}  \\
&& \simeq \frac{a^3}{4}\ln^2(1-u_1)\, \sum _{n=-\infty}^\infty (-1)^n
\int _{-\infty}^\infty \frac{d\nu }{\nu ^2+\frac{n^2}{4}}\,
\left(w^*\right)^{i\nu -\frac{n}{2}}\,
\left(w\right)^{i\nu +\frac{n}{2}}\,E^2_{\nu,n}, \nonumber
\eeqn
where we used the fact that in the Regge limit $1-u_1\simeq (\mathbf{k}_1+\mathbf{k}_2)^2/s_2$.

The integral in RHS of Eq.~\ref{delta3} can be easily   obtained using the calculations of the previous section. Namely, we have shown that
Eq.~\ref{NLOR} gives the $\nu,n$ representation of the NLLA contribution to the remainder function at two loops, namely
\beqn\label{stam1}
\frac{1}{2}\sum_{n=-\infty}^{\infty}\int d\nu \frac{(-1)^n}{\nu^2+\frac{n^2}{4}}\left(E^2_{\nu,n}-\frac{1}{4}\frac{n^2}{ \left(\nu^2+\frac{n^2}{4}\right)^2}
\right)\left(w^*\right)^{i\nu -\frac{n}{2}}\,
\left(w\right)^{i\nu +\frac{n}{2}}.
\eeqn

We calculate in separate the transform of the second  term in the brackets in Eq.~\ref{stam1}
\beqn\label{stam555}
\sum_{n=-\infty}^{\infty}\int d\nu \frac{(-1)^n}{\nu^2+\frac{n^2}{4}}\left(-\frac{1}{4}\frac{n^2}{ \left(\nu^2+\frac{n^2}{4}\right)^2} \right)\left(w^*\right)^{i\nu -\frac{n}{2}}\,
\left(w\right)^{i\nu +\frac{n}{2}}.
\eeqn
The integral in Eq.~\ref{stam555} can be calculated  using the residue theorem closing the contour either in the upper semiplane for poles $\nu=i|n|/2$ and multiplying the residue by $i2\pi$, or in the lower semiplane for poles $\nu=-i|n|/2$ and then multiplying the residue by $-i2\pi$.
The result has $w \leftrightarrow 1/w$ symmetry so that it is enough to consider only contributions for $|w|<1$.
The residue at $\nu=-i|n|/2$ for $|w|<1$ reads
\beqn
&&-i2\pi\text{Res}\left(\frac{(-1)^n}{\nu^2+\frac{n^2}{4}}\frac{1}{2}\left(-\frac{1}{4}\frac{n^2}{ \left(\nu^2+\frac{n^2}{4}\right)^2} \right)\left(w^*\right)^{i\nu -\frac{n}{2}}\,
\left(w\right)^{i\nu +\frac{n}{2}},-i|n|/2\right)\\
&&=-\frac{3}{2}\frac{ (-1)^n \pi  ({w^*})^{|n|}}{|n|^3}+\frac{3 (-1)^n \pi  ({w^*})^{|n|} \ln |w|^2}{4 n^2}-\frac{(-1)^n \pi  ({w^*})^{|n|} \ln^2 |w|^2 }{8 |n|}.
\nonumber
\eeqn
The summation over $n$ (for $n>0$) is readily performed using the series representation of the polylogarithms $\text{Li}_n(x)=\sum_{k=1}^{\infty}x^k/k^n$
and we get
\beqn\label{resstam1}
&&-i2\pi\sum_{n=1}^{\infty}\text{Res}\left(\frac{(-1)^n}{\nu^2+\frac{n^2}{4}}\frac{1}{2}\left(-\frac{1}{4}\frac{n^2}{ \left(\nu^2+\frac{n^2}{4}\right)^2} \right)\left(w^*\right)^{i\nu -\frac{n}{2}}\,
\left(w\right)^{i\nu +\frac{n}{2}},-i|n|/2\right)\\
&& =\frac{1}{8} \pi  \ln^2|w|^2 \ln(1+w^*)+\frac{3}{4} \pi \ln |w|^2 \text{Li}_2(-w^*)-\frac{3}{2} \pi  \text{Li}_3(-w^*).\nonumber
\eeqn
The contribution from the sum over negative $n$ is  added by substitution $w^* \to w$ and we obtain
\beqn\label{stam33}
&&\sum_{n=-\infty}^{\infty}\int d\nu \frac{(-1)^n}{\nu^2+\frac{n^2}{4}}\frac{1}{2}\left(-\frac{1}{4}\frac{n^2}{ \left(\nu^2+\frac{n^2}{4}\right)^2} \right)\left(w^*\right)^{i\nu -\frac{n}{2}}\,
\left(w\right)^{i\nu +\frac{n}{2}}\\
&& =\frac{1}{8} \pi  \ln^2|w|^2 \ln|1+w|^2+\frac{3}{4} \pi \ln |w|^2 \left(\text{Li}_2(-w)-\text{Li}_2(-w^*)\right)-\frac{3}{2} \pi  \text{Li}_3(-w)
-\frac{3}{2} \pi  \text{Li}_3(-w^*). \nonumber
\eeqn

From Eq.~\ref{delta3} it follows that for the  three-loop LLA contribution we need to calculate the following expression
\beqn\label{stam66}
\sum_{n=-\infty}^{\infty}\int d\nu \frac{(-1)^n}{\nu^2+\frac{n^2}{4}}E^2_{\nu,n}\left(w^*\right)^{i\nu -\frac{n}{2}}\,
\left(w\right)^{i\nu +\frac{n}{2}}.
\eeqn
This can be obtained by subtracting Eq.~\ref{stam33} from Eq.~\ref{stam1}. In the previous section we found that
\beqn\label{stam11}
&&\frac{1}{2}\sum_{n=-\infty}^{\infty}\int d\nu \frac{(-1)^n}{\nu^2+\frac{n^2}{4}}\left(E^2_{\nu,n}-\frac{1}{4}\frac{n^2}{ \left(\nu^2+\frac{n^2}{4}\right)^2}
\right)\left(w^*\right)^{i\nu -\frac{n}{2}}\,
\left(w\right)^{i\nu +\frac{n}{2}}\\
&&= \frac{\pi}{2}\ln|w|^2 \ln^2|1+w|^2-\frac{\pi}{3}\ln^3|1+w|^2+\pi \ln|w|^2 \left(\text{Li}_2(-w)+\text{Li}_2(-w^*)\right) \nonumber\\
&&
\hspace{1cm}-2\pi \left(\text{Li}_3(-w)+\text{Li}_3(-w^*)\right) \nonumber
\eeqn
and thus we write
\beqn\label{stam666}
&& \frac{1}{2}\sum_{n=-\infty}^{\infty}\int d\nu \frac{(-1)^n}{\nu^2+\frac{n^2}{4}}E^2_{\nu,n}\left(w^*\right)^{i\nu -\frac{n}{2}}\,
\left(w\right)^{i\nu +\frac{n}{2}}= \frac{\pi}{2}\ln|w|^2 \ln^2|1+w|^2 -\frac{\pi}{3}\ln^3|1+w|^2 \hspace{1cm}\;\;\;
\\
&&-\frac{\pi}{8}\ln^2|w|^2\ln|1+w|^2+\frac{\pi}{4} \ln|w|^2 \left(\text{Li}_2(-w)+\text{Li}_2(-w^*)\right)
 -\frac{\pi}{2} \left(\text{Li}_3(-w)+\text{Li}_3(-w^*)\right). \nonumber
\eeqn
Finally from Eq.~\ref{delta3} and Eq.~\ref{stam666} we obtain the remainder function at three loops in the leading logarithm approximation~(LLA), namely
\beqn\label{R63app}
&& a^3 R_{6}^{(3)\;LLA}=i\Delta^{(3)} _{2\rightarrow 4}=i\pi \frac{a^3}{4} \ln^2(1-u_1)\left(
 \ln|w|^2\ln^2|1+w|^2-\frac{2}{3}\ln^3|1+w|^2\right. \hspace{1cm}\;\;\;
\\
&&\left.-\frac{1}{4}\ln^2|w|^2 \ln|1+w|^2+\frac{1}{2} \ln|w|^2 \left(\text{Li}_2(-w)+\text{Li}_2(-w^*)\right)
 - \text{Li}_3(-w)-\text{Li}_3(-w^*)\right). \nonumber
\eeqn
The complex variables $w$  is expressed in terms of the reduced cross ratios  of Eq.~\ref{redcross} as
 \beqn
w=\frac{1-z}{z}=\frac{B^+}{\tilde{u}_2},\;\;\; w^*=\frac{1-z^*}{z^*}=\frac{B^-}{\tilde{u}_2}
 \eeqn
for $B^{\pm}$ defined in Eq.~\ref{Bpm} by
\begin{eqnarray}
B^{\pm}=\frac{1-\tilde{u}_2-\tilde{u}_3\pm \sqrt{(1-\tilde{u}_2-\tilde{u}_3)^2-4\tilde{u}_2\tilde{u}_3}}{2}.
\end{eqnarray}


\setcounter{equation}{0}

\renewcommand{\theequation}{H.\arabic{equation}}

\section{The real part of the remainder function at three loops in  NLLA}\label{app:real3NLLA}

In this section we calculate the real part of the remainder function at three loops in the next-to-leading logarithmic approximation~(NLLA). The expression for $\Re(R^{(3)NLLA}_6)$ is obtained expanding the dispersion relation Eq.~\ref{disp} in powers of the perturbation expansion parameter~$a$. It is worth emphasizing that the calculation of $\Re(R^{(3)NLLA}_6)$ does not require the knowledge of  currently unavailable  subleading corrections to the BFKL eigenvalue $\omega(\nu,n)$.
We plug the LLA function $f^{LLA}(\omega)$ of Eq.~\ref{fomegaLLA} in Eq.~\ref{disp}  and expand it in $a$ to the third order
\beqn\label{expdispapp}
i\pi \delta a^2 R^{(2)}_6 +a^3 R^{(3)}_6-\frac{i\pi \delta^3}{6}\simeq \frac{ia^3}{4}(\ln(1-u_1)+i\pi)^2\sum_{n=-\infty}^{\infty} \int d\nu \frac{(-1)^n}{\nu^2+\frac{n^2}{4}}E^2_{\nu,n}\left(w^*\right)^{i\nu -\frac{n}{2}}\,
\left(w\right)^{i\nu +\frac{n}{2}},\;\;\;
\eeqn
where $w$ and $E_{\nu,n}$ are given by Eq.~\ref{wapp} and Eq.~\ref{Enun} respectively. The phases $\delta$ and $\omega_{ab}$  of Eq.~\ref{deltaomega} can be written as
\beqn
\delta=\frac{a}{2}\ln (\tilde{u}_2 \tilde{u}_3)=\frac{a}{2}\ln \frac{|w|^2}{|1+w|^4},\;\; \omega_{ab}=\frac{a}{2}\ln \frac{\tilde{u}_3} {\tilde{u}_2}=\frac{a}{2}\ln|w|^2
\eeqn
using the leading order term for the cusp anomalous dimension $\gamma_{K}\simeq 4 a$.
The equation Eq.~\ref{expdispapp} is valid only for the LLA term and the real part of the NLLA term of the remainder function at three loops.
Solving it for the LLA term we get
\beqn
R^{(3)\; LLA}_6=\frac{i}{4} \ln^2(1-u_1)\sum_{n=-\infty}^{\infty} \int d\nu \frac{(-1)^n}{\nu^2+\frac{n^2}{4}} \left(w^*\right)^{i\nu -\frac{n}{2}}\,
\left(w\right)^{i\nu +\frac{n}{2}}E^2_{\nu,n}
\eeqn
in full agreement with Eq.~\ref{delta3}.

Next we solve Eq.~\ref{expdispapp} for the real part of the NLLA remainder function
\beqn\label{reNLLAeq}
&& \hspace{-1cm} \Re(R^{(3)\;NLLA}_6)=-\frac{i\pi \delta R^{(2)\;LLA}_6}{a}
-\frac{\pi}{2} \ln(1-u_1)\sum_{n=-\infty}^{\infty} \int d\nu \frac{(-1)^n E^2_{\nu,n}}{\nu^2+\frac{n^2}{4}} \left(w^*\right)^{i\nu -\frac{n}{2}}\,
\left(w\right)^{i\nu +\frac{n}{2}}\;\;\;\nonumber\\
&&=-\frac{i\pi \delta R^{(2)\;LLA}_6}{a}-\frac{2\pi R^{(3)\; LLA}_6}{i\ln(1-u_1)}.
\eeqn
From Eq.~\ref{reNLLAeq} we see that the next-to-leading logarithmic contribution is related to the leading logarithmic terms at two and three loops.
The function $R^{(2)\;LLA}$ was found using BFKL approach in ref.~\cite{BLS2} and  given by the first term on RHS of Eq.~\ref{R62w}
\beqn
R^{(2)\;LLA}_6=\frac{i \pi}{2}\ln(1-u_1)\ln |1+w|^2 \ln\left|1+\frac{1}{w}\right|^2.
\eeqn
The LLA remainder function at three loops $R^{(3)\;LLA}$ was calculated in the appendix~\ref{app:3loops} and is given by Eq.~\ref{R63app}.
Summing up all terms in Eq.~\ref{reNLLAeq} we readily obtain
\beqn\label{R63NLLAapp}
&&  \Re(R^{(3)\;NLLA}_6)= \frac{\pi^2}{4} \ln (1-u_1)\left(
 \ln|w|^2\ln^2|1+w|^2-\frac{2}{3}\ln^3|1+w|^2\right. \hspace{1cm}\;\;\;
\\
&&\left.-\frac{1}{2}\ln^2|w|^2 \ln|1+w|^2- \ln|w|^2 \left(\text{Li}_2(-w)+\text{Li}_2(-w^*)\right)
 +2 \text{Li}_3(-w)+2\text{Li}_3(-w^*)\right). \nonumber
\eeqn
Note that $\Re(R^{(3)\;NLLA}_6)$ resembles very much the form of $R^{(3)\;LLA}$ in Eq.~\ref{R63app} as one could expect from Eq.~\ref{reNLLAeq}.
The complex variables $w$  is expressed in terms of the reduced cross ratios  of Eq.~\ref{redcross} as
 \beqn
w=\frac{1-z}{z}=\frac{B^+}{\tilde{u}_2},\;\;\; w^*=\frac{1-z^*}{z^*}=\frac{B^-}{\tilde{u}_2}
 \eeqn
for $B^{\pm}$ defined in Eq.~\ref{Bpm} by
\begin{eqnarray}
B^{\pm}=\frac{1-\tilde{u}_2-\tilde{u}_3\pm \sqrt{(1-\tilde{u}_2-\tilde{u}_3)^2-4\tilde{u}_2\tilde{u}_3}}{2}.
\end{eqnarray}

\end{document}